\RequirePackage{fix-cm}
\documentclass[smallextended]{svjour3}       % onecolumn (second format)

\smartqed  % flush right qed marks, e.g. at end of proof
\usepackage{graphicx}
\usepackage{cite,amssymb,amsmath,amsfonts,graphicx,bm,dcolumn,mathrsfs}
\usepackage{amsmath,amssymb,amsfonts,graphicx,bm,dcolumn,mathrsfs}

\newcommand{\dd}{\mbox{d}}
\newcommand{\ii}{\mathrm{i}}

\begin{document}

\title{Computation of microcanonical entropy at fixed magnetization without direct counting}

\author{Alessandro Campa$^1$, Giacomo Gori$^{2,7}$, Vahan Hovhannisyan$^3$, Stefano Ruffo$^{4,5}$ and Andrea Trombettoni$^{6,7,4}$}

\institute{Alessandro Campa \at
alessandro.campa@iss.it
           \and
           Giacomo Gori \at
              gori@sissa.it
              \and
           Vahan Hovhannisyan \at
              v.hovhannisyan@yerphi.am
              \and
           Stefano Ruffo \at
             ruffo@sissa.it
              \and
           Andrea Trombettoni \at
              andreatr@sissa.it
              \and
              $^1$ National Center for Radiation Protection and
Computational Physics, Istituto Superiore di Sanit\`{a},
Viale Regina Elena 299, 00161 Roma, Italy  \at
\and
$^2$ Institut f\"ur Theoretische Physik, Universit\"at Heidelberg, D-69120 Heidelberg, Germany \at
\and
$^3$ A. I. Alikhanyan National Science Laboratory, 2 Alikhanian Brothers St., 0036 Yerevan, Armenia
\at
\and
$^4$ SISSA, via Bonomea 265, I-34136 Trieste, Italy \&
INFN, Sezione di Trieste, I-34151 Trieste, Italy
\at
\and
$^5$ Istituto dei Sistemi Complessi, CNR, via Madonna del Piano 10,
I-50019 Sesto Fiorentino, Italy \at
\and
$^6$ Department of Physics, University of Trieste, Strada Costiera 11, I-34151 Trieste, Italy \at
\and
$^7$ CNR-IOM DEMOCRITOS Simulation Center, Via Bonomea 265, I-34136 Trieste, Italy
}

\date{Received: \today}

\authorrunning{A. Campa et al.}

\titlerunning{Microcanonical entropy at fixed magnetization}

\maketitle

\begin{abstract}
We discuss a method to compute the microcanonical entropy at fixed magnetization without direct counting. Our approach is based
on the evaluation of a saddle-point leading to an optimization problem. The method is applied to a benchmark Ising model with
simultaneous presence of mean-field and nearest-neighbour interactions for which direct counting is indeed possible, thus
allowing a comparison. Moreover, we apply the method to an Ising model with mean-field, nearest-neighbour and next-nearest-neighbour
interactions, for which direct counting is not straightforward. For this model, we compare the solution obtained by our method
with the one obtained from the formula for the entropy in terms of all correlation functions. This example shows that for general couplings
our method is much more convenient than direct counting methods to compute the microcanonical entropy at fixed magnetization.

\end{abstract}

\section{Introduction}
\label{intro}
The determination of the entropy of a physical system is a major task in any thermodynamic calculation \cite{Huang}. To compute the
entropy, as notoriously carved on the Boltzmann tombstone, one has to compute the number of microscopic states consistent with the
macroscopic quantities characterizing the system. The central problem is then the counting of the number of such states. When the
system is simple, by means of combinatoric tools one can perform explicitly this calculation for any finite number $N$ of the
constituents of the system. In other cases one does not have the access to the exact number at finite $N$, but the correct limit
for large $N$ can be found by neglecting subleading contributions. We refer to the possibility of directly determining the number
of states -- in an exact or approximate form -- as {\it direct counting} or enumeration, as it is also referred to in \cite{Pathria}.
We stress that the problem of the difficulty of computing the entropy by direct counting is particularly relevant when dealing
with non-additive systems, like systems with long-range interactions, since these systems present, most of the times, ensemble
inequivalence, and thus the microcanical entropy is in general different from the canonical entropy \cite{PR}.

When direct counting is not easy or possible, due to the difficulty of carrying out the corresponding full or approximate
combinatoric calculation, one can anyway aim at calculating the entropy in the limit of large $N$ resorting to other accessible
thermodynamic quantities or using information from the equation of state of the system under consideration. A problem that is often
encountered is represented by the difficulty of getting the needed expressions in presence of additional conserved quantities,
consequence of specific physical constraints acting on the system.
This issue is present, e.g., when, working in the microcanonical ensemble, the entropy at a given energy has to be
determined in presence of other constraints. In this case the entropy, and therefore the number of
states, has to be known as a function not only of the energy, but also of other macroscopic observables, such as order parameters
and other correlation functions, characterizing the thermodynamic state.

Among the various physical systems in which the previous general considerations apply, a special attention is devoted to magnetic
and spin systems, where the issue of determining the entropy from the counting of states is ubiquitously present \cite{Parisi}. The
order parameter characterizing these systems is the magnetization. Then there are situations in which it could be necessary to compute
the entropy as a function of the energy and magnetization, or, in other words, to find the entropy in presence of the constraint of
fixed magnetization. Generally the natural physical setting of a magnetic system is one in which the fixed control parameter is the
external magnetic field, with the (average) magnetization obtained as a derived quantity by the usual thermodynamic relation; also
the spontaneous magnetization in absence of an external field in ferromagnetic systems falls into this scheme. However, there are
cases in which the Hamiltonian of a magnetic system can be used to describe another type of system, meaning that it is possible to
map the original system into a magnetic one; the magnetization would then represent, under this mapping, a control parameter of the
original system. For example, the spin value could be associated to the presence or absence of a charged particle in the corresponding
site, and the magnetization would be related to the density of the system \cite{Misawa}. It is clear that in such cases the
magnetization is a natural control parameter. When ensembles are equivalent, the computation of the entropy could be done in the
canonical ensemble, which is generally easier to deal with. But if equivalence does not hold, as in most non-additive systems, in
particular those with long-range interactions, if one wants to study the system at given energy (and magnetization), it is necessary
to compute the microcanonical partition function, i.e., to compute the number of states. In this circumstance, the interest of
this computation is present also when the Hamiltonian represents a genuine magnetic system; in fact, inequivalence would be associated
to a possible negative susceptibility in the microcanonical ensemble. To determine if this property occurs, one has to compute the
microcanonical entropy at fixed magnetization.

Another typical situation in which one works at a fixed magnetization in magnetic systems is provided by spin models obtained in the
strong coupling limit from lattice fermionic or bosonic systems, such as the Hubbard model \cite{Auerbach} or its bosonic
counterpart \cite{Matsubara}. In that case, for large interactions one naturally obtains spin models \cite{Lacroix} since in that limit
the part of the Hilbert space contributing to the effective Hamiltonian for each lattice site is finite dimensional. When the number of
particles in the original lattice model is fixed, the magnetization in the spin model is in turn fixed and therefore one is interested
to work in sectors at a fixed magnetization.

Since the calculation of entropy at fixed magnetization without resorting to direct counting is in general a challenging problem, in this
paper we aim at presenting a method that, starting from the \textit{canonical} partition function, is able to give a useful expression
for the microcanonical entropy at fixed magnetization. As a benchmark, we first apply the presented method to a model in which there is
a long-range, mean-field coupling between all the spins of an Ising chain in presence of a nearest-neighbour term
\cite{Nagle70,Kardar83,Mukamel05}. The rationale for this choice is that it may exhibit ensemble inequivalence when the long-range
coupling is ferromagnetic and the short-range is antiferromagnetic, and one can study and compare both the canonical and microcanonical
phase diagrams. Even more importantly for our purposes, in this model the direct counting is possible and one can compare the results
obtained from the method presented here and direct counting findings. Moreover, the model has the merit that it is easily generalizable,
and one can study the effect of additional couplings, such as finite-range terms. In a recent paper \cite{Campa19} the addition of a
next-nearest-neighbour term was considered and the canonical phase diagram shown to exhibit a rich structure, with a large
variety of different critical points. Since the direct counting in such a model is rather involved and cumbersome, it provides an ideal
case study to give results for the microcanonical entropy at fixed magnetization in a case in which direct counting is not known in
the literature.

The plan of the paper is the following. In section 2 we derive formal expressions for the canonical and microcanonical partition
functions at fixed magnetization. In section 3 these expressions are used to obtain the procedure to compute the canonical and
microcanonical entropies at fixed magnetization for a general spin system in which there are both long-range mean-field and short-range
interactions. In this section we also discuss the issue of ensemble inequivalence. In section 4 we apply the procedure to specific
models; in particular we consider a model where the short-range interaction is only between nearest-neighbours and another model where
also a next-nearest-neighbour interaction is present. We choose these models since they allow (with difficulty for the second case)
a direct counting evaluation, and thus a comparison with the results of our procedure provides a test for it. In section 5 a discussion
and the conclusions are given. Some additional material is in the appendices.

\section{Formal expressions for the partition functions at fixed magnetization}
\label{sec:1}

In this paper we consider spin systems, for which the dynamical variables take discrete values; correspondingly, the sum over the
configurations is denoted by a sum over the discrete values of the spins $S_i$. The constraint of fixed magnetization is a constraint
on the sum of the $S_i$. However, the general expressions that we obtain in this section and in the following one are independent from
the structure of the configuration space of the system. In particular, they are valid also for continuous dynamical variables (in that
case the constraint would be, e.g., on the position of the center of mass of the system): in the derivation of the general expressions
one would substitute the sum over the spin configurations $\{S_i\}$ with the integral over the continuous dynamical variables, obtaining
at the end the same expressions.

As explained above, we focus on the entropy at fixed magnetization, which is defined by:
\begin{equation}
\label{magdef}
\hat{m} = \frac{1}{N} \sum_{i=1}^N S_i \, .
\end{equation}
The use of the hat in the notation is justified by the necessity to distinguish the magnetization $\hat{m}$ as a function of the spin
configuration from its fixed value $m$ on which the thermodynamic quantities defined in the following will depend. Before proceeding,
we find it convenient to begin by writing down the known expressions for the usual entropy and the free energy, i.e., for the case when
there is no such constraint of fixed magnetization. In a system of $N$ spins, the number of states (i.e., the microcanonical partition
function) $\Omega(\epsilon,N)$ with fixed energy per particle equal to $\epsilon$ and the associated microcanonical entropy
$s_{{\rm micr}}(\epsilon)$ are given by:
\begin{equation}
\Omega(\epsilon,N) \equiv \exp \left[ N s_{{\rm micr}}(\epsilon) \right] =
\sum_{\{S_i\}} \delta \left( N\epsilon - H(\{S_i\}) \right) \, .
\label{micrentrgen}
\end{equation}
On the other hand, the partition function $Z(\beta,N)$ and the associated (rescaled) free energy $\phi(\beta)$ per particle are
expressed by:
\begin{eqnarray}
Z(\beta,N) &\equiv& \exp \left[ - N \phi (\beta) \right] =
\sum_{\{S_i\}} \exp \left[ -\beta H(\{S_i\}) \right] \nonumber \\
&=& \int \dd (N\epsilon) \exp \left\{ -N\left[ \beta \epsilon - s_{{\rm micr}}(\epsilon) \right] \right\} \, ,
\label{canfreegen}
\end{eqnarray}
where $\beta=1/k_B T$ is proportional to the inverse of the temperature $T$, with $k_B$ the Boltzmann constant.
The last expression shows that in the thermodynamic limit, where the saddle-point approximation becomes exact, the rescaled free energy
$\phi(\beta)$ is the Legendre-Fenchel transform of the microcanonical entropy $s_{{\rm micr}}(\epsilon)$:
\begin{equation}
\phi(\beta) = \min_{\epsilon} \left[ \beta \epsilon - s_{{\rm micr}}(\epsilon) \right] \, .
\label{lftransfgen}
\end{equation}

To consider now the magnetization constraint, we need to modify the above expressions by adding a proper $\delta$ function. Precisely,
the number of states (or microcanonical partition function) $\widetilde{\Omega}(\epsilon,m,N)$ with fixed energy per particle equal to
$\epsilon$ and fixed magnetization per particle equal to $m$, and the associated microcanonical entropy per particle
$\widetilde{s}_{{\rm micr}}(\epsilon,m)$ are obtained by:
\begin{equation}
\widetilde{\Omega}(\epsilon,m,N) \equiv \exp \left[ N \widetilde{s}_{{\rm micr}}(\epsilon,m) \right] =
\sum_{\{S_i\}} \delta \left( N\epsilon - H(\{S_i\}) \right) \, \delta \left( \sum_i S_i -Nm \right) \, .
\label{micrpart}
\end{equation}
Analogously, the partition function $\widetilde{Z}(\beta,m,N)$ and the associated rescaled free energy $\widetilde{\phi}(\beta,m)$ per
particle at fixed magnetization $m$ are given by:
\begin{eqnarray}
\widetilde{Z}(\beta,m,N) &\equiv& \exp \left[ - N \widetilde{\phi} (\beta,m) \right] =
\sum_{\{S_i\}} \exp \left[ -\beta H(\{S_i\}) \right] \delta \left( \sum_i S_i -Nm \right) \nonumber \\
&=& \int \dd (N\epsilon) \exp \left\{ -N\left[ \beta \epsilon - \widetilde{s}_{{\rm micr}}(\epsilon,m) \right] \right\} \, .
\label{canpart}
\end{eqnarray}
As in the case of Eq. (\ref{lftransfgen}), the last expression shows that in the thermodynamic limit the rescaled free energy
$\widetilde{\phi}(\beta,m)$ is the Legendre-Fenchel transform of the microcanonical entropy $\widetilde{s}_{{\rm micr}}(\epsilon,m)$:
\begin{equation}
\widetilde{\phi} (\beta,m) = \min_{\epsilon} \left[ \beta \epsilon - \widetilde{s}_{{\rm micr}}(\epsilon,m) \right] \, ,
\label{lftransf}
\end{equation}
(for brevity in the following we do not specify any more that the entropy and the free energy are to be intended per particle).

The above expressions can be transformed by using the representation of the $\delta$ function. Since the partition functions are
given by sums over spin configurations that assume discrete values, in principle the $\delta$ functions should not be interpreted as
Dirac $\delta$, but as Kronecker $\delta$. This should be reflected in the corresponding representation. However, the distinction
between Dirac and Kronecker $\delta$ is not relevant for the final expression, as will be clear in the following (besides, in the
thermodynamic limit both $\epsilon$ and $m$ become parameters assuming continuous values). We can therefore use the representation
of the Dirac $\delta$ function.
For the microcanonical partition function we thus have:
\begin{eqnarray}
\label{micrpartspin}
&&\,\,\,\,\,\,\,\,\,\,\,\,\,\,\,\,\,\,\,\,\,\,\,\,\,\,\,\,\,\,\, \widetilde{\Omega}(\epsilon,m,N) = \\
&=& \sum_{\{S_i\}} \left( \frac{1}{2\pi}\right)^2 \!\!\!\! \int_{-\infty}^{+\infty} \!\!\!\!\!\! \dd \lambda
\int_{-\infty}^{+\infty} \!\!\!\!\!\! \dd \varphi \, \exp \left\{ \ii \lambda \left[ N\epsilon - H(\{S_i\})\right]
+\ii \varphi \left[ \sum_i S_i -Nm \right] \right\} \nonumber \\
&=& \sum_{\{S_i\}} \left( \frac{1}{2\pi \ii}\right)^2 \!\!\!\! \int_{-\ii \infty}^{+\ii \infty} \!\!\!\!\!\! \dd \lambda
\int_{-\ii \infty}^{+\ii \infty} \!\!\!\!\!\! \dd \varphi \, \exp \left\{ \lambda \left[ N\epsilon - H(\{S_i\})\right]
+\varphi \left[ \sum_i S_i -Nm \right] \right\}. \nonumber
\end{eqnarray}
For the canonical partition function we similarly have:
\begin{equation}
\widetilde{Z}(\beta,m,N) = \sum_{\{S_i\}} \frac{1}{2\pi \ii} \int_{-\ii \infty}^{+\ii \infty} \!\!\!\!\!\! \dd \varphi
\exp \left\{ - \beta H(\{S_i\}) +\varphi \left[ \sum_i S_i -Nm \right] \right\} \, .
\label{canpartspin}
\end{equation}
The microcanonical entropy and the rescaled free energy, as shown in the defining equivalences at the beginning of Eq. (\ref{micrpart})
and Eq. (\ref{canpart}), are then obtained from the logarithm of Eq. (\ref{micrpartspin}) and Eq. (\ref{canpartspin}), respectively.
We note that in both expressions (\ref{micrpartspin}) and (\ref{canpartspin}) the sum over the configurations is the canonical
partition function of the system with an added external magnetic field, in which $\varphi$ is equal to the magnetic field multiplied
by the inverse temperature.

In the next section we apply these general expressions to the case where long-range interactions are present.

\section{Models with long- and short-range terms}
\label{sec:1bis}

We are interested in models that can exhibit ensemble inequivalence, arising from the presence of long-range interactions. In this
framework and considering spin systems, the latter are defined as those in which the coupling constant between two spins has a decaying
behaviour with distance as $J_{i,j} \sim 1/|i-j|^\alpha$, with $\alpha$ smaller than the spatial dimension of the system. When
$\alpha = 0$ we have the case of mean-field terms. When long-range interactions are present, thermodynamic quantities are no more
additive and ensemble inequivalence can arise. In particular, this generally occurs in presence of first order phase transitions in
the canonical ensemble, so that the function $\phi(\beta)$ is not everywhere differentiable; it is known, in fact, that if
$\phi(\beta)$ is everywhere differentiable, then Eq. (\ref{lftransfgen}) can be inverted \cite{PR}, so that $s(\epsilon)$ is concave
and ensembles are equivalent\footnote{We remind that the Legendre-Fenchel transform of any function is automatically concave.}.

To have a concrete case of study, we consider models having general finite-range interactions plus long-range terms of the mean-field
form. These models generally present first order phase transitions, due to the combined effect of the mean-field terms and of the
short-range terms. We will derive our expressions by assuming a system with two mean-field terms plus unspecified short-range
interactions. The dimensionality $d$ of the lattice will also be left unspecified. Of course, as in any computational method, the
computations in concrete models are easier for one-dimensional systems. In the following section, where we show the application to a
specific model, we will then consider, as an example, a one-dimensional model that we have already studied within the framework of the
canonical ensemble \cite{Campa19} and that, even with only one mean-field term, presents a very rich thermodynamic phase diagrams,
with first and second order phase transitions, critical and tricritical points, and critical end points.

As a preliminary step, we recall that, dealing with systems with long-range interactions, one often obtains an expression of the
canonical partition function $Z(\beta,N)$ in the form
\begin{equation}
Z(\beta,N) = \int_{-\infty}^{+\infty} \dd \mathbf{x} \exp \left[ - N \psi(\beta,\mathbf{x}) \right] \, ,
\label{zgenmult}
\end{equation}
where $\mathbf{x} \equiv (x_1,\dots,x_M)$ is a $M$-dimensional auxiliary variable, and where $\psi(\beta,\mathbf{x})$ is a real
analytic function of $\beta$ and $\mathbf{x}$. This form is a multidimensional generalization \cite{campa04} of an expression
previously considered only for $M=1$ \cite{leyv2002}. As we will see shortly, one obtains an expression of this sort by using a
Hubbard-Stratonovich transformation to perform the computation of the canonical partition function in presence of mean-field terms.
Using a saddle-point evaluation, valid in the thermodynamic limit, one finds that the microcanonical entropy
$s_{{\rm micr}}(\epsilon)$ is given by \cite{PR,campa04,leyv2002}:
\begin{equation}
s_{{\rm micr}}(\epsilon) = \max_{\mathbf{x}} \left\{ \min_{\beta} \left[ \beta \epsilon - \psi(\beta,\mathbf{x}) \right] \right\} \, .
\label{smicgen}
\end{equation}
On the other hand, the canonical entropy, computed from the rescaled free energy $\phi(\beta)$, is obtained from
\begin{equation}
s_{{\rm can}}(\epsilon) = \min_{\beta} \left\{ \max_{\mathbf{x}} \left[ \beta \epsilon - \psi(\beta,\mathbf{x}) \right] \right\} \, .
\label{scangen}
\end{equation}
These two min-max expressions can give different results \cite{PR,campa04,leyv2002}, and when this happens ensemble inequivalence occurs.

In this section we will consider the case with two auxiliary variables (i.e., $M=2$), thus an expression of the form
\begin{equation}
Z(\beta,N) = \int_{-\infty}^{+\infty} \!\!\!\!\!\! \dd x \int_{-\infty}^{+\infty} \!\!\!\!\!\! \dd y
\exp \left[ - N \psi(\beta,x,y) \right] \, .
\label{zgen}
\end{equation}
Let us then introduce the kind of models we consider. As mentioned above, we will work with a system with two mean-field terms, more
precisely with a Hamiltonian given by:
\begin{equation}
H(\{S_i\}) = - \frac{J}{2N} \left( \sum_{i=1}^N S_i \right)^2 - \frac{K}{2N} \left( \sum_{i=1}^N S_i^2 \right)^2
+ \sum_{i=1}^N U([S_i]) \, ,
\label{modhamext}
\end{equation}
with positive coupling constants, $J>0$, $K>0$. We see that the first mean-field term is proportional to $-N\hat{m}^2$, with the
magnetization $\hat{m}$ defined in Eq. (\ref{magdef}), while the second mean-field term is proportional to $-N\hat{q}^2$, with
$\hat{q}$ equal to the quadrupole moment
\begin{equation}
\label{quaddef}
\hat{q} = \frac{1}{N} \sum_{i=1}^N S_i^2 \, .
\end{equation}
As usual with mean-field terms, the coupling constants are normalized with the number $N$ of spins. In the final term of the Hamiltonian
the notation $U([S_i])$ denotes a function of the $i$-th spin $S_i$ and its neighbours. Namely, $U([S_i])$ is a short-range term that
could contain the interaction of the $i$-th spin with its nearest-neighbours, next-nearest-neighbours, next-to-next-nearest-neighbours,
and so on. If the lattice is not one-dimensional, of course the index $i$ stands for the set of indices used to identify the lattice
point. In the implementation of the method to an one-dimensional lattice in section \ref{secmodels} we consider in detail Ising spins
with $U([S_i])=-(K_1/2) S_i S_{i+1}$ (only nearest-neighbour couplings) and $U([S_i])=-(K_1/2) S_i S_{i+1}-(K_2/2) S_i S_{i+2}$
(including a next-nearest-neighbour term). However, the general expressions that we will derive are independent on the spin value. As
a matter of fact, in order to have a nontrivial contribution to the Hamiltonian in correspondence of the quadrupole mean-field term,
one has to consider non Ising spins\footnote{\label{footbeg}For $S_i= -1,0,1$ and a function $U([S_i])$ given by just a term
proportional to $S_i^2$
we would obtain the Hamiltonian of the Blume-Emery-Griffiths model; however, in this computation we are not assuming a specific spin
model and a specific function $U([S_i])$.}.

We now make use of the Hubbard-Stratonovich transformation
\begin{equation}
\label{hubtran}
\exp (ab^2) = \sqrt{\frac{a}{\pi} }\int_{-\infty}^{+\infty} \!\!\!\!\!\! \dd x \, \exp (-ax^2 +2abx)
\,\,\,\,\,\,\,\,\,\,\,\,\,\,\,\,\,\, (a > 0) \, ,
\end{equation}
applied, for positive $\beta$, once to the case where $a=\beta J N/2$ and $b=\left(\sum_i S_i /N\right) = \hat{m}$, and once to the
case where $a=\beta K N/2$ and $b=\left(\sum_i S_i^2 /N\right) = \hat{q}$. We then obtain:
\begin{eqnarray}
\label{hubb2}
&&\,\,\,\,\,\,\,\,\,\,\,\,\,\,\,\,\,\,\,\,\,\,\,\,\,\,\,\,\,\,\, \exp \left[ -\beta H(\{S_i\}) \right] = \\
&=& \frac{\beta N \sqrt{JK}}{2\pi} \int_{-\infty}^{+\infty} \!\!\!\!\!\! \dd x
\int_{-\infty}^{+\infty} \!\!\!\!\!\! \dd y \exp \Big[ -\frac{N}{2}\beta J x^2 -\frac{N}{2}\beta K y^2 \nonumber \\
&& + \beta J x \sum_{i=1}^N S_i + \beta K y \sum_{i=1}^N S_i^2 - \beta \sum_{i=1}^N U([S_i]) \Big] \, . \nonumber
\end{eqnarray}
Inserting in Eq. (\ref{canpartspin}) and performing the sum over the spin configurations we get
\begin{eqnarray}
\label{canpart0two}
&&\,\,\,\,\,\,\,\,\,\,\,\,\,\,\,\,\,\,\,\,\,\,\,\,\,\,\,\,\,\,\, \widetilde{Z}(\beta,m,N) = \\
&=& \frac{1}{2\pi \ii} \frac{\beta N \sqrt{JK}}{2\pi} \int_{-\infty}^{+\infty} \!\!\!\!\!\! \dd x
\int_{-\infty}^{+\infty} \!\!\!\!\!\! \dd y \int_{-\ii \infty}^{+\ii \infty} \!\!\!\!\!\! \dd \varphi
\exp \Big\{ -N \Big[ \frac{\beta J}{2} x^2 + \frac{\beta K}{2} y^2 \nonumber \\
&&+ \hat{\psi}(\beta,\beta J x + \varphi,\beta K y) + m\varphi \Big] \Big\} \, , \nonumber
\end{eqnarray}
where the function $\hat{\psi}(\beta,\beta J x + \varphi,\beta K y)$ is defined by:
\begin{eqnarray}
\label{psihatdeftwo}
&&\,\,\,\,\,\,\,\,\,\,\,\,\,\,\,\,\,\,\,\,\,\,\,\,\,\,\,\,\,\,\,
\exp \left[ - N \hat{\psi}(\beta,\beta J x + \varphi,\beta K y) \right] = \\
&=& \sum_{\{S_i\}} \exp \left[ \left( \beta J x + \varphi \right) \sum_{i=1}^N S_i + \beta K y \sum_{i=1}^N S_i^2
- \beta \sum_{i=1}^N U([S_i]) \right] \, . \nonumber
\end{eqnarray}
Then, we have an expression of the canonical partition function $Z(\beta,m,N)$ in which, besides the two auxiliary variables $x$ and
$y$, the other auxiliary variable $\varphi$, coming from the Fourier representation of the $\delta$ function that implements the
constraint of fixed magnetization $m$, appears. We see that $\hat{\psi}$ is a real analytic function of its arguments.
Thus, the dependence of $\hat{\psi}$ on $\beta J x + \varphi$ comes from the combination of the third term in the exponent of
Eq. (\ref{hubb2}) and the second term in the exponent of Eq. (\ref{canpartspin}), the dependence on $\beta K y$ comes from the fourth
term in the exponent of Eq. (\ref{hubb2}), while the extra dependence on $\beta$ comes from the short-range term, the last term in
the exponent of Eq. (\ref{hubb2}). It might be difficult to obtain $\hat{\psi}$, for example the use of a transfer matrix evaluation
could be necessary. Here we do not specify any particular form. Clearly, if $d>1$ the determination of $\hat{\psi}$ becomes
considerably more involved (one may resort to approximations for it), but for the purposes of the present discussion there are no
changes in the argument. We remark that these difficulties would be present already in the calculation of the canonical partition
function. Moreover, so far we did not explicitly use the values taken by the spins $S_i$. As remarked above, in the implementation
we will consider Ising spins, $S_i=\pm 1$, but the application to more general cases is straightforward.

Furthermore, for the moment we consider only positive temperatures, i.e. $\beta \ge 0$, postponing the treatment of negative
temperatures. In spin systems, where the energy is upper bounded, the latter are possible in the microcanonical ensemble. They occur
for energies where the derivative of the entropy with respect to the energy is negative. Although physically we do not envisage a
thermal bath at negative temperatures, it is possible to formally define a canonical partition function at negative temperatures,
since the upper boundedness of the energy implies that this partition function is well defined. We remind that, thinking to the
thermodynamic situations in which it is sensible to talk of negative temperatures, we are forced to consider them as ``hotter'' than
the positive temperatures. More precisely, $T = +\infty$ and $T = -\infty$ coincide, while any negative temperature is ``hotter''
than $T = \infty$. Moreover,  if $T_1 < T_2 < 0$, then $T_2$ is ``hotter'' than $T_1$. Finally, the ``hottest'' temperature is
$T = 0^-$, although numerically it is infinitesimally close to the coldest temperature $T = 0^+$.

Before writing the expression for the microcanonical partition function (\ref{micrpartspin}) we remark the following. In
(\ref{micrpartspin}) the integrals on $\lambda$ and $\varphi$ are made on the imaginary axis; however, we can perform the integration
on a line parallel to the imaginary axis, adding a real part to both $\lambda$ and $\varphi$, since this is allowed by the definition
of the Dirac $\delta$. Furthermore, as discussed in \cite{PR}, the integrals in $\lambda$ and $\varphi$ can be limited to a finite
segment parallel to the imaginary axis\footnote{This also shows why, as noted above, the use of the representation of the Dirac $\delta$,
instead of that of the Kronecker $\delta$, has no importance in this computation.}.

We then obtain:
\begin{eqnarray}
\label{micrpart0two}
&&\,\,\,\,\,\,\,\,\,\,\,\,\,\,\,\,\,\,\,\,\,\,\,\,\,\,\,\,\,\,\, \widetilde{\Omega}(\epsilon,m,N) =
\left( \frac{1}{2\pi \ii}\right)^2 \frac{\beta N \sqrt{JK}}{2\pi} \times \\
&\times& \int_{-\infty}^{+\infty} \!\!\!\!\!\! \dd x \int_{-\infty}^{+\infty} \!\!\!\!\!\! \dd y
\int_{\sigma -\ii \eta}^{\sigma +\ii \eta} \!\!\!\!\!\! \dd \lambda \int_{\mu -\ii \nu}^{\mu +\ii \nu} \!\!\!\!\!\! \dd \varphi
\exp \Big\{ N \Big[ \lambda\epsilon - \frac{\lambda J}{2} x^2 - \frac{\beta K}{2} y^2 \nonumber \\
&& - \hat{\psi}(\lambda,\lambda J x + \varphi,\lambda K y) - m\varphi \Big] \Big\} \, , \nonumber
\end{eqnarray}
where $\sigma$ and $\mu$ are the fixed real parts of $\lambda$ and $\varphi$, respectively, while $\eta$ and $\nu$ denote the respective
limits of integration along the segments parallel to the imaginary axis. Analogously, the integration limits of $\varphi$ in
Eq. (\ref{canpart0two}) can be changed in $(\mu - \ii \nu,\mu + \ii \nu)$. We note that, since Eq. (\ref{hubb2}) is valid for
positive $\beta$, then the fixed real part $\sigma$ in the integral in $\lambda$ must be nonnegative.

From Eqs. (\ref{canpart0two}) and (\ref{micrpart0two}) one can derive the expressions giving the rescaled free energy
$\widetilde{\phi}(\beta,m)$ and the microcanonical entropy $\widetilde{s}(\epsilon,m)$. The integrals can be evaluated with the saddle
point approximation. It can be shown \cite{PR} that the relevant saddle points in the variables $\lambda$ and $\varphi$ lie on the
real axis. Let us first consider $\widetilde{\phi}(\beta,m)$. Since the real part of the exponent in Eq. (\ref{canpart0two}) has a
minimum on the real axis when $\varphi$ varies on the line parallel to the imaginary axis, then it has a maximum, on the same point
of the real axis, when $\varphi$ varies along the real axis. We therefore have:
\begin{equation}
\widetilde{\phi}(\beta,m) = \min_{x,y} \left[ \max_{\varphi} \left( \frac{\beta J}{2}x^2 + \frac{\beta K}{2} y^2
+ \hat{\psi}(\beta,\beta J x +\varphi,\beta K y) + m \varphi \right) \right] \, .
\label{exphitwo}
\end{equation}
From this one can obtain the expression for the canonical entropy $\widetilde{s}_{{\rm can}}(\epsilon,m)$
\begin{eqnarray}
\label{canentrtwo}
&&\,\,\,\,\,\,\,\,\,\,\,\,\,\,\,\,\,\,\,\,\,\,\,\,\,\,\,\,\,\,\, \widetilde{s}_{{\rm can}}(\epsilon,m) = \\
&=& \min_{{\beta \ge 0}} \left\{ \max_{x,y} \left[ \min_{\varphi} \left(
\beta \epsilon -\frac{\beta J}{2}x^2 - \frac{\beta K}{2} y^2
- \hat{\psi}(\beta,\beta J x + \varphi,\beta K y) - m \varphi \right) \right] \right \} \, . \nonumber
\end{eqnarray}
An analogous saddle point evaluation of Eq. (\ref{micrpart0two}) allows to find the expression of the microcanonical
entropy $\widetilde{s}_{{\rm micr}}(\epsilon,m)$, obtaining:
\begin{eqnarray}
\label{micrentrtwo}
&&\,\,\,\,\,\,\,\,\,\,\,\,\,\,\,\,\,\,\,\,\,\,\,\,\,\,\,\,\,\,\, \widetilde{s}_{{\rm micr}}(\epsilon,m) = \\
&=& \max_{x,y} \left\{ \min_{{\beta \ge 0}} \left[ \min_{\varphi} \left(
\beta \epsilon -\frac{\beta J}{2}x^2 - \frac{\beta K}{2} y^2
- \hat{\psi}(\beta,\beta J x + \varphi,\beta K y) - m \varphi \right) \right] \right \} \, . \nonumber
\end{eqnarray}
From Eqs. (\ref{canentrtwo}) and (\ref{micrentrtwo}) one can obtain the $(\beta,x,y,\varphi)$ point satisfying the extremal problems.
As in the case of $s_{{\rm micr}}(\epsilon)$ and $s_{{\rm can}}(\epsilon)$, given respectively in Eq. (\ref{smicgen}) and
Eq. (\ref{scangen}), the different order in which minimization with respect to $\beta$ and maximization with respect to $x$ and $y$
is performed can lead to different results, i.e., to different extremal points, and then to ensemble inequivalence.
From the properties of min-max extremal problems \cite{PR} it follows that in general we will have
$\widetilde{s}_{{\rm micr}}(\epsilon,m) \le \widetilde{s}_{{\rm can}}(\epsilon,m)$.

We note that in Eqs. (\ref{canentrtwo}) and (\ref{micrentrtwo}) it is possible to include $\beta = 0$. In fact, for $\beta = 0$
the partition function (\ref{canpartspin}) reduces to the number of states with given magnetization, and no Hubbard-Stratonovich
transformation is necessary. However, the function in round brackets in Eqs. (\ref{canentrtwo}) and (\ref{micrentrtwo}) becomes
equal to $(-\hat{\psi}(0,\varphi,0) - m\varphi)$, and it is easy to see that this value\footnote{It is also not difficult to realize
from Eq. (\ref{psihatdeftwo}) that $\exp \left[ -N \hat{\psi}(0,\varphi) \right]$ is equal to the partition function of $N$ independent
spins subject to a magnetic field $h$ with $\varphi$ playing the role of $\beta h$.} is reached continuously when $\beta \to 0^+$.
Therefore, the two extremal problems (\ref{canentrtwo}) and (\ref{micrentrtwo}) can be extended to $\beta = 0$, for which no
maximization with respect to $x$ and $y$ has to be performed.

The study of the extremal problems (\ref{canentrtwo}) and (\ref{micrentrtwo}) is performed by determining the stationarity and
stability conditions that have to be satisfied by the $(\beta,x,y,\varphi)$ point in each case. We emphasize that the relations we
are going to derive have the purpose to find analytical expressions for the points, but in an actual computation concerning a given
concrete model the most rapid way to proceed will be to numerically solve the extremization problems (\ref{canentrtwo}) and
(\ref{micrentrtwo}). Therefore, the somewhat cumbersome appearance of the expressions that we will obtain are not a hindrance for the
applications. To ease the notation it is convenient to denote with $u$, $v$ and $w$ the three arguments
($\beta$, $\beta J x +\varphi$ and $\beta K y$) of $\hat{\psi}$, and, as customary, to use subscripts to denote partial derivatives
with respect to an argument.

We will proceed step by step for each of the two extremal problems, since this is convenient to determine the stability conditions,
expressed by inequalities to be satisfied at the extremal points $(\beta,x,y,\varphi)$.
On the other hand, the stationarity conditions can be easily written all together, and they are the same for both problems, and we
anticipate them here. They are given by:
\begin{eqnarray}
\label{cond1}
\hat{\psi}_v + m &=& 0 \\
\label{cond2}
\hat{\psi}_v + x &=& 0 \\
\label{cond3}
\hat{\psi}_w + y &=& 0 \\
\label{cond4}
\epsilon -\frac{J}{2}x^2 -\hat{\psi}_u - J x\hat{\psi}_v - K y\hat{\psi}_w &=& 0 \, .
\end{eqnarray}
The first three equations are also those that must be verified in the extremal problem (\ref{exphitwo}). The first two equations imply
that at the extremal points we have $x=m$. The fact that at the extremal point one has $x=m$ is consistent with what obtained in the
study of the unconstrained problem, i.e., in the computation of $\phi(\beta)$ and $s(\epsilon)$ (canonical or microcanonical), where
one derives the equilibrium magnetization and finds that it is equal to the extremal value of $x$ \cite{PR}.

Let us know complete the analysis, obtaining for each one of the stationarity conditions summarized in Eqs. (\ref{cond1}-\ref{cond4})
the corresponding stability condition. We begin with the problem (\ref{canentrtwo}). Minimizing with respect to $\varphi$ one has:
\begin{eqnarray}
\label{stattwo1}
\hat{\psi}_v + m &=& 0 \\
\label{stabtwo1}
\hat{\psi}_{vv} &<& 0 \, .
\end{eqnarray}
Eq. (\ref{stattwo1}) gives $\varphi$ as a function of $(\beta,x,y,m)$, and one obtains the following relations:
\begin{eqnarray}
\label{varphitwox}
\varphi_x &=& -\beta J\\
\label{varphitwoy}
\varphi_y &=& -\beta K \frac{\hat{\psi}_{vw}}{\hat{\psi}_{vv}}\\
\label{varphitwob}
\varphi_{\beta} &=& -\frac{\hat{\psi}_{uv}}{\hat{\psi}_{vv}} - J x - Ky \frac{\hat{\psi}_{vw}}{\hat{\psi}_{vv}} \, ,
\end{eqnarray}
useful for the successive steps. Then we have now:
\begin{eqnarray}
\label{canentrtwo_a}
&& \widetilde{s}_{{\rm can}}(\epsilon,m) = \min_{{\beta \ge 0}} \Big\{ \max_{x,y} \Big[
\beta \epsilon -\frac{\beta J}{2}x^2 - \frac{\beta K}{2} y^2 \nonumber \\
&&- \hat{\psi}(\beta,\beta J x + \varphi(\beta,x,y,m),\beta K y) - m \varphi(\beta,x,y,m) \Big] \Big\}  \, .
\end{eqnarray}
The maximization with respect to $x$ and $y$ leads to the following stationarity and stability conditions:
\begin{eqnarray}
\label{stattwo2x}
\hat{\psi}_v + x &=& 0 \\
\label{stattwo2y}
\hat{\psi}_w + y &=& 0 \\
1 + \beta K \left( \hat{\psi}_{ww} - \frac{\hat{\psi}_{vw}^2}{\hat{\psi}_{vv}} \right) &>& 0 \, ,
\label{stabtwo2}
\end{eqnarray}
where use has been made of Eqs. (\ref{varphitwox}) and (\ref{varphitwoy}). Equation (\ref{stabtwo2}) will be compared with the
corresponding one obtained below in the study of $\widetilde{s}_{{\rm micr}}(\epsilon,m)$, to see how inequivalence can arise.
Eqs. (\ref{stattwo2x}) and (\ref{stattwo2y}) define in principle $x$ and $y$ as a function of $(\beta,m)$. However, the former one,
taken together with Eq. (\ref{stattwo1}), shows that at the extremum one has $x=m$. Consistently, computing the derivative of
$x$ and $y$ with respect to $\beta$, used in the following step, we find, with the help of (\ref{varphitwob}):
\begin{eqnarray}
\label{xtwobeta}
x_{\beta} &=& 0 \\
\label{ytwobeta}
y_{\beta} &=& - \frac{\hat{\psi}_{uw} - \frac{\hat{\psi}_{uv}\hat{\psi}_{vw}}{\hat{\psi}_{vv}} + Ky
\left( \hat{\psi}_{ww} - \frac{\hat{\psi}_{vw}^2}{\hat{\psi}_{vv}} \right)}
{1 + \beta K \left( \hat{\psi}_{ww} - \frac{\hat{\psi}_{vw}^2}{\hat{\psi}_{vv}} \right)} \, .
\end{eqnarray}
Then we are left with:
\begin{eqnarray}
\label{canentrtwo_b}
&& \widetilde{s}_{{\rm can}}(\epsilon,m) = \min_{{\beta \ge 0}} \Big\{
\beta \epsilon -\frac{\beta J}{2}x^2(\beta,m) - \frac{\beta K}{2} y^2(\beta,m) \nonumber \\
&&- \hat{\psi}(\beta,\beta J x(\beta,m) + \varphi(\beta,x(\beta,m),y(\beta,m),m),\beta K y(\beta,m)) \nonumber \\
&&- m \varphi(\beta,x(\beta,m),y(\beta,m),m) \Big\}  \, ,
\end{eqnarray}
where in this expression we have left the formal dependence of $x$ and $y$ on $(\beta,m)$. Without writing explicitly anymore this
dependence, minimization with respect to $\beta$ leads to the stationarity condition:
\begin{equation}
\label{stattwo3}
\epsilon - \frac{J}{2}x^2 - \frac{K}{2}y^2 - \hat{\psi}_u - Jx\hat{\psi}_v - Ky\hat{\psi}_w = 0 \, ,
\end{equation}
where use has been made of Eqs. (\ref{stattwo1}), (\ref{xtwobeta}) and (\ref{ytwobeta}). This equation gives $\beta$ as a function
of $\epsilon$ and $m$. The stability condition requires some algebra. Using Eq. (\ref{varphitwob}) the stability is obtained as:
\begin{eqnarray}
&&\hat{\psi}_{uu}{\hat{\psi}_{vv}} - \hat{\psi}_{uv}^2
+ 2 Ky \left( \hat{\psi}_{uw}{\hat{\psi}_{vv}} - \hat{\psi}_{uv}\hat{\psi}_{vw} \right)
+ \left(Ky\right)^2 \left( {\hat{\psi}_{vv}}\hat{\psi}_{ww} - \hat{\psi}_{vw}^2 \right) \nonumber \\
&+&\beta K \hat{\psi}_{vv} \left[ \hat{\psi}_{uw} - \frac{\hat{\psi}_{uv}\hat{\psi}_{vw}}{\hat{\psi}_{vv}} + Ky
\left( \hat{\psi}_{ww} - \frac{\hat{\psi}_{vw}^2}{\hat{\psi}_{vv}} \right) \right]y_{\beta} > 0 \, .
\label{stabtwo3}
\end{eqnarray}
Substituting the expression of $y_{\beta}$ given by Eq. (\ref{ytwobeta}) we have:
\begin{eqnarray}
\label{stabtwo3b}
&&\hat{\psi}_{uu}{\hat{\psi}_{vv}} - \hat{\psi}_{uv}^2
+ 2 Ky \left( \hat{\psi}_{uw}{\hat{\psi}_{vv}} - \hat{\psi}_{uv}\hat{\psi}_{vw} \right)
+ \left(Ky\right)^2 \left( {\hat{\psi}_{vv}}\hat{\psi}_{ww} - \hat{\psi}_{vw}^2 \right) \nonumber \\
&-&\beta K \hat{\psi}_{vv} \frac{\left[ \hat{\psi}_{uw} - \frac{\hat{\psi}_{uv}\hat{\psi}_{vw}}{\hat{\psi}_{vv}} + Ky
\left( \hat{\psi}_{ww} - \frac{\hat{\psi}_{vw}^2}{\hat{\psi}_{vv}} \right) \right]^2}
{1 + \beta K \left( \hat{\psi}_{ww} - \frac{\hat{\psi}_{vw}^2}{\hat{\psi}_{vv}} \right)} > 0 \, .
\end{eqnarray}
From Eq. (\ref{stabtwo1}) and Eq. (\ref{stabtwo2}) it follows that the second line of Eq. (\ref{stabtwo3b}) (including the minus sign)
is positive. This is to be taken into account in the comparison with the corresponding stability condition in the following study
of $\widetilde{s}_{{\rm micr}}(\epsilon,m)$.

It is convenient to summarize the stability conditions of the problem (\ref{canentrtwo}). They are:
\begin{eqnarray}
\label{canstab1}
&&\hat{\psi}_{vv} < 0 \\
\label{canstab2}
&&1 + \beta K \left( \hat{\psi}_{ww} - \frac{\hat{\psi}_{vw}^2}{\hat{\psi}_{vv}} \right) > 0 \\
&&\hat{\psi}_{uu}{\hat{\psi}_{vv}} - \hat{\psi}_{uv}^2
+ 2 Ky \left( \hat{\psi}_{uw}{\hat{\psi}_{vv}} - \hat{\psi}_{uv}\hat{\psi}_{vw} \right)
+ \left(Ky\right)^2 \left( {\hat{\psi}_{vv}}\hat{\psi}_{ww} - \hat{\psi}_{vw}^2 \right)  \nonumber \\
&&-\beta K \hat{\psi}_{vv} \frac{\left[ \hat{\psi}_{uw} - \frac{\hat{\psi}_{uv}\hat{\psi}_{vw}}{\hat{\psi}_{vv}} + Ky
\left( \hat{\psi}_{ww} - \frac{\hat{\psi}_{vw}^2}{\hat{\psi}_{vv}} \right) \right]^2}
{1 + \beta K \left( \hat{\psi}_{ww} - \frac{\hat{\psi}_{vw}^2}{\hat{\psi}_{vv}} \right)} > 0 \, .
\label{canstab3}
\end{eqnarray}

We now consider the extremal problem (\ref{micrentrtwo}), concerning the microcanonical entropy. The first step is the same of the
canonical case, so that the corresponding stationarity and stability conditions are given by (\ref{stattwo1}) and (\ref{stabtwo1}),
respectively, and also Eqs. (\ref{varphitwox}), (\ref{varphitwoy}) and (\ref{varphitwob}) are the same.
Then we have:
\begin{eqnarray}
\label{micrentrtwo_a}
&& \widetilde{s}_{{\rm micr}}(\epsilon,m) = \max_{x,y} \Big\{ \min_{{\beta \ge 0}} \Big[
\beta \epsilon -\frac{\beta J}{2}x^2 - \frac{\beta K}{2} y^2 \nonumber \\
&&- \hat{\psi}(\beta,\beta J x + \varphi(\beta,x,y,m),\beta K y) - m \varphi(\beta,x,y,m) \Big] \Big\}  \, .
\end{eqnarray}
The minimization with respect to $\beta$ gives the stationarity condition:
\begin{equation}
\label{stattwo4}
\epsilon - \frac{J}{2}x^2 - \frac{K}{2}y^2 - \hat{\psi}_u - Jx\hat{\psi}_v - Ky\hat{\psi}_w = 0 \, ,
\end{equation}
where Eq. (\ref{stattwo1}) has been used. As we know, it is the same as that of the other extremal problem. It gives $\beta$ as a
function of $(\epsilon,x,y,m)$. From this function we obtain, using (\ref{varphitwob}):
\begin{eqnarray}
\label{betatwox}
\beta_x &=& 0 \\
\label{betatwoy}
\beta_y &=& -\beta K \frac{\hat{\psi}_{uw} - \frac{\hat{\psi}_{uv}\hat{\psi}_{vw}}{\hat{\psi}_{vv}} + Ky
\left( \hat{\psi}_{ww} - \frac{\hat{\psi}_{vw}^2}{\hat{\psi}_{vv}}\right)}
{\hat{\psi}_{uu} - \frac{\hat{\psi}_{uv}^2}{\hat{\psi}_{vv}}
+ 2 Ky \left( \hat{\psi}_{uw} - \frac{\hat{\psi}_{uv}\hat{\psi}_{vw}}{\hat{\psi}_{vv}} \right)
+ \left(Ky\right)^2 \left( \hat{\psi}_{ww} - \frac{\hat{\psi}_{vw}^2}{\hat{\psi}_{vv}} \right)} \, .
\end{eqnarray}
We point out that in writing the last expressions we have also used the stationarity conditions that are obtained in the following
and last step, the maximization with respect to $x$ and $y$. Making use of Eq. (\ref{varphitwob}), we find that the stability condition
is given by:
\begin{equation}
\label{stabtwo4}
\hat{\psi}_{uu}{\hat{\psi}_{vv}} - \hat{\psi}_{uv}^2
+ 2 Ky \left( \hat{\psi}_{uw}{\hat{\psi}_{vv}} - \hat{\psi}_{uv}\hat{\psi}_{vw} \right)
+ \left(Ky\right)^2 \left( {\hat{\psi}_{vv}}\hat{\psi}_{ww} - \hat{\psi}_{vw}^2 \right) > 0 \, .
\end{equation}
We see that this is different from the corresponding stability condition (\ref{stabtwo3b}). The latter requires the positivity of an
expression given by the left hand side of (\ref{stabtwo4}) plus the second line of (\ref{stabtwo3b}), that we have noted is always
positive at the extremal point. We come back to this later. The final step is given by:
\begin{eqnarray}
\label{micrentrtwo_b}
&& \widetilde{s}_{{\rm micr}}(\epsilon,m) = \max_{x,y} \Big\{
\beta(\epsilon,x,y,m) \epsilon -\frac{\beta(\epsilon,x,y,m) J}{2}x^2 - \frac{\beta(\epsilon,x,y,m) K}{2} y^2 \nonumber \\
&&- \hat{\psi}(\beta(\epsilon,x,y,m),\beta(\epsilon,x,y,m) J x + \varphi(\beta(\epsilon,x,y,m),x,y,m),\beta(\epsilon,x,y,m) K y) \nonumber \\
&&- m \varphi(\beta(\epsilon,x,y,m),x,y,m) \Big\}  \, .
\end{eqnarray}
The stationarity conditions are:
\begin{eqnarray}
\label{stattwo5x}
\hat{\psi}_v + x &=& 0 \\
\label{stattwo5y}
\hat{\psi}_w + y &=& 0 \, ,
\end{eqnarray}
equal respectively to (\ref{stattwo2x}) and (\ref{stattwo2y}), as expected. But the stability condition is not equal to (\ref{stabtwo2}),
being instead given by, using (\ref{varphitwox}) and (\ref{varphitwoy}):
\begin{eqnarray}
\label{stabtwo5}
&&1 + \beta K \left[ \hat{\psi}_{ww} - \frac{\hat{\psi}_{vw}^2}{\hat{\psi}_{vv}} \right] \\
&+&\left[ \hat{\psi}_{uw} - \frac{\hat{\psi}_{uv}\hat{\psi}_{vw}}{\hat{\psi}_{vv}} + Ky
\left( \hat{\psi}_{ww} - \frac{\hat{\psi}_{vw}^2}{\hat{\psi}_{vv}}\right) \right] \beta_y > 0 \, . \nonumber
\end{eqnarray}
Substituting $\beta_y$ from eq. (\ref{betatwoy}) we obtain:
\begin{eqnarray}
\label{stabtwo5b}
&&1 + \beta K \left[ \hat{\psi}_{ww} - \frac{\hat{\psi}_{vw}^2}{\hat{\psi}_{vv}} \right] \\
&-&\beta K \frac{\left[ \hat{\psi}_{uw} - \frac{\hat{\psi}_{uv}\hat{\psi}_{vw}}{\hat{\psi}_{vv}} + Ky
\left( \hat{\psi}_{ww} - \frac{\hat{\psi}_{vw}^2}{\hat{\psi}_{vv}}\right) \right]^2}
{\hat{\psi}_{uu} - \frac{\hat{\psi}_{uv}^2}{\hat{\psi}_{vv}}
+ 2 Ky \left( \hat{\psi}_{uw} - \frac{\hat{\psi}_{uv}\hat{\psi}_{vw}}{\hat{\psi}_{vv}} \right)
+ \left(Ky\right)^2 \left( \hat{\psi}_{ww} - \frac{\hat{\psi}_{vw}^2}{\hat{\psi}_{vv}} \right)} > 0 \, . \nonumber
\end{eqnarray}
Furthermore, from Eqs. (\ref{stabtwo1}) and (\ref{stabtwo4}) we have that the denominator in the fraction in the second line is negative;
therefore the second line (including the minus sign) is positive. On the other hand, we had noted, in the study of the canonical problem,
that the corresponding stability condition (see Eq. (\ref{stabtwo2})) required that the first line alone be positive. After treating
the case of negative temperatures we come back to the differences between the stability conditions and the associated possibility of
ensemble inequivalence.

The summary of the stability conditions of the problem (\ref{micrentrtwo}) is:
\begin{eqnarray}
\label{micrstab1}
&&\hat{\psi}_{vv} < 0 \\
\label{micrstab2}
&&\hat{\psi}_{uu}{\hat{\psi}_{vv}} - \hat{\psi}_{uv}^2
+ 2 Ky \left( \hat{\psi}_{uw}{\hat{\psi}_{vv}} - \hat{\psi}_{uv}\hat{\psi}_{vw} \right) \nonumber \\
&&+ \left(Ky\right)^2 \left( {\hat{\psi}_{vv}}\hat{\psi}_{ww} - \hat{\psi}_{vw}^2 \right) > 0 \\
\label{micrstab3}
&&1 + \beta K \left[ \hat{\psi}_{ww} - \frac{\hat{\psi}_{vw}^2}{\hat{\psi}_{vv}} \right]  \\
&&-\beta K \frac{\left[ \hat{\psi}_{uw} - \frac{\hat{\psi}_{uv}\hat{\psi}_{vw}}{\hat{\psi}_{vv}} + Ky
\left( \hat{\psi}_{ww} - \frac{\hat{\psi}_{vw}^2}{\hat{\psi}_{vv}}\right) \right]^2}
{\hat{\psi}_{uu} - \frac{\hat{\psi}_{uv}^2}{\hat{\psi}_{vv}}
+ 2 Ky \left( \hat{\psi}_{uw} - \frac{\hat{\psi}_{uv}\hat{\psi}_{vw}}{\hat{\psi}_{vv}} \right)
+ \left(Ky\right)^2 \left( \hat{\psi}_{ww} - \frac{\hat{\psi}_{vw}^2}{\hat{\psi}_{vv}} \right)}  > 0 \, . \nonumber
\end{eqnarray}

\subsection{Negative temperatures}
\label{secnegtemp}

We noted above that spin systems can have negative temperatures, since the energy is upper bounded. As a consequence, we should
expect that, if in Eqs. (\ref{canentrtwo}) or (\ref{micrentrtwo}) we choose values of $\epsilon$ and $m$ for which the corresponding
temperature is negative, then the extremal problems will not be satisfied for any $\beta \ge 0$. Then we have to extend the analysis
to negative values of $\beta$. The treatment of negative temperatures requires some changes in the expressions. However, we will
see below that, considering the two cases together, we can obtain a procedure that has the double advantage to be shorter and to
include at the same time temperatures of both signs.

When $\beta < 0$ we have to use a different form of the Hubbard-Stratonovich transformation, i.e.
\begin{equation}
\label{hubtran_neg}
\exp (ab^2) = \sqrt{\frac{-a}{\pi} }\int_{-\infty}^{+\infty} \!\!\!\!\!\! \dd x \, \exp (ax^2 +2\ii abx) \, ,
\,\,\,\,\,\,\,\,\,\,\,\,\,\,\,\,\,\,
\end{equation}
which is valid for $a<0$. For the following analysis it is useful to note that this equality is valid also if we add fixed imaginary
parts to $x$ and to $y$ , i.e., if we perform the $x$ and $y$ integral on a line parallel to the real $x$ axis and the real $y$ axis,
respectively. The expression for the canonical partition function that replaces (\ref{canpart0two}) is then
\begin{eqnarray}
\label{canpart0twoneg}
&&\,\,\,\,\,\,\,\,\,\,\,\,\,\,\,\,\,\,\,\,\,\,\,\,\,\,\,\,\,\,\, \widetilde{Z}(\beta,m,N) = \\
&=& \frac{1}{2\pi \ii} \frac{|\beta| N \sqrt{JK}}{2\pi} \int_{-\infty}^{+\infty} \!\!\!\!\!\! \dd x
\int_{-\infty}^{+\infty} \!\!\!\!\!\! \dd y \int_{\mu -\ii \nu}^{\mu +\ii \nu} \!\!\!\!\!\! \dd \varphi
\exp \Big\{ -N \Big[ -\frac{\beta J}{2} x^2 - \frac{\beta K}{2} y^2 \nonumber \\
&&+ \hat{\psi}(\beta,\ii \beta J x + \varphi,\ii \beta K y) + m\varphi \Big] \Big\} \, , \nonumber
\end{eqnarray}
where we have already taken into account that the integral over $\varphi$ can be on a line parallel to the imaginary axis, with real
part equal to $\mu$, and that the integration limits of $\varphi$ can be changed in $(\mu - \ii \nu,\mu + \ii \nu)$.
The function $\hat{\psi}(\beta,\ii \beta J x + \varphi,\ii \beta K y)$ is defined as in Eq. (\ref{psihatdeftwo}), therefore by the
right hand side of that expression with $\beta$ substituted by $\ii \beta$. In the same way, the expression of the microcanonical
partition function replacing (\ref{micrpart0two}) is
\begin{eqnarray}
\label{micrpart0twoneg}
&&\,\,\,\,\,\,\,\,\,\,\,\,\,\,\,\,\,\,\,\,\,\,\,\,\,\,\,\,\,\,\, \widetilde{\Omega}(\epsilon,m,N) =
\left( \frac{1}{2\pi \ii}\right)^2 \frac{|\beta| N \sqrt{JK}}{2\pi} \times \\
&\times& \int_{-\infty}^{+\infty} \!\!\!\!\!\! \dd x \int_{-\infty}^{+\infty} \!\!\!\!\!\! \dd y
\int_{\sigma -\ii \eta}^{\sigma +\ii \eta} \!\!\!\!\!\! \dd \lambda \int_{\mu -\ii \nu}^{\mu +\ii \nu} \!\!\!\!\!\! \dd \varphi
\exp \Big\{ N \Big[ \lambda\epsilon + \frac{\lambda J}{2} x^2 + \frac{\beta K}{2} y^2 \nonumber \\
&& - \hat{\psi}(\lambda,\ii \lambda J x + \varphi,\ii \lambda K y) - m\varphi \Big] \Big\} \, , \nonumber
\end{eqnarray}
where $\sigma$ is the fixed real parts of $\lambda$, and we have taken into account that the integration limits of $\lambda$ can be
taken as $(\sigma - \ii \eta,\sigma + \ii \eta)$. We note that, since Eq. (\ref{hubtran_neg}) is valid for negative $\beta$, then the
fixed real part $\sigma$ in the integral in $\lambda$ must be nonpositive.

Then, we obtain the following expressions replacing Eqs. (\ref{exphitwo}), (\ref{canentrtwo}) and (\ref{micrentrtwo}):
\begin{equation}
\widetilde{\phi}(\beta,m) = \min_{x,y} \left[ \max_{\varphi} \left( -\frac{\beta J}{2}x^2 - \frac{\beta K}{2} y^2
+ \hat{\psi}(\beta,\ii \beta J x +\varphi,\ii \beta K y) + m \varphi \right) \right] \, ,
\label{exphitwoneg}
\end{equation}
for the rescaled free energy,
\begin{eqnarray}
\label{canentrtwoneg}
&&\,\,\,\,\,\,\,\,\,\,\,\,\,\,\,\,\,\,\,\,\,\,\,\,\,\,\,\,\,\,\, \widetilde{s}_{{\rm can}}(\epsilon,m) = \\
&=& \min_{{\beta \le 0}} \left\{ \max_{x,y} \left[ \min_{\varphi} \left(
\beta \epsilon +\frac{\beta J}{2}x^2 + \frac{\beta K}{2} y^2
- \hat{\psi}(\beta,\ii \beta J x + \varphi,\ii \beta K y) - m \varphi \right) \right] \right \} \, , \nonumber
\end{eqnarray}
for the canonical entropy, and
\begin{eqnarray}
\label{micrentrtwoneg}
&&\,\,\,\,\,\,\,\,\,\,\,\,\,\,\,\,\,\,\,\,\,\,\,\,\,\,\,\,\,\,\, \widetilde{s}_{{\rm micr}}(\epsilon,m) = \\
&=& \max_{x,y} \left\{ \min_{{\beta \le 0}} \left[ \min_{\varphi} \left(
\beta \epsilon +\frac{\beta J}{2}x^2 + \frac{\beta K}{2} y^2
- \hat{\psi}(\beta,\ii \beta J x + \varphi,\ii \beta K y) - m \varphi \right) \right] \right \} \, , \nonumber
\end{eqnarray}
for the microcanonical entropy. For the same argument given before, the value $\beta = 0$ can be included in the analysis. Depending
on the values of $\epsilon$ and $m$ we expect that it is possible to satisfy either the extremal problem (\ref{canentrtwo}) or the
extremal problem (\ref{canentrtwoneg}), but not both (except when they are both satisfied for $\beta = 0$); the same for the couple
of problems (\ref{micrentrtwo}) and (\ref{micrentrtwoneg}).

We can now follow the same steps as above to obtain the stationarity and stability conditions of the problems (\ref{canentrtwoneg})
and (\ref{micrentrtwoneg}). We will make a shorter presentation than for positive $\beta$, in particular we will not mention explicitly
the equations used to obtain the stability conditions. In both cases the first step, i.e., the minimization with respect to $\varphi$,
is the same as before. We then have the same stationarity and stability conditions, namely
\begin{eqnarray}
\label{phi1neg}
\hat{\psi}_v + m &=& 0 \\
\label{phi2neg}
\hat{\psi}_{vv} &<& 0  \, .
\end{eqnarray}
In writing, here and in the following, an inequality like (\ref{phi2neg}) for a quantity that in principle is complex, we are assuming
that it is actually real. In fact, as it can be verified a posteriori, both $\ii x$ and $\ii y$ at the extremal points are real
quantities. Eq. (\ref{phi1neg}), defining $\varphi$ as a function of $(\beta,x,y,m)$, now gives:
\begin{eqnarray}
\label{varphitwoxneg}
\varphi_x &=& -\ii \beta J\\
\label{varphitwoyneg}
\varphi_y &=& -\ii \beta K \frac{\hat{\psi}_{vw}}{\hat{\psi}_{vv}}\\
\label{varphitwobneg}
\varphi_{\beta} &=& -\frac{\hat{\psi}_{uv}}{\hat{\psi}_{vv}} - \ii J x - \ii Ky \frac{\hat{\psi}_{vw}}{\hat{\psi}_{vv}} \, .
\end{eqnarray}
Thus, in the second step for the canonical case we have
\begin{eqnarray}
\label{canentrtwo_aneg}
&& \widetilde{s}_{{\rm can}}(\epsilon,m) = \min_{{\beta \le 0}} \Big\{ \max_{x,y} \Big[
\beta \epsilon +\frac{\beta J}{2}x^2 + \frac{\beta K}{2} y^2 \nonumber \\
&&- \hat{\psi}(\beta,\ii \beta J x + \varphi(\beta,x,y,m),\ii \beta K y) - m \varphi(\beta,x,y,m) \Big] \Big\} \, .
\end{eqnarray}
The maximization with respect to $x$ and $y$ leads to the following stationarity and stability conditions:
\begin{eqnarray}
\label{stattwo2xneg}
-\ii \hat{\psi}_v + x &=& 0 \\
\label{stattwo2yneg}
-\ii \hat{\psi}_w + y &=& 0 \\
1 + \beta K \left( \hat{\psi}_{ww} - \frac{\hat{\psi}_{vw}^2}{\hat{\psi}_{vv}} \right)&>& 0 \, .
\label{stabtwo2neg}
\end{eqnarray}
This stationarity condition (\ref{stattwo2xneg}), defining $x$ as a function of $(\beta,m)$, shows, together with Eq. (\ref{phi1neg}),
that $\ii x=m$, that we substitute in the next and final step. Then, as before, $x_{\beta} = 0$. We note that the application of the
saddle point requires that the integral in $x$ be performed on a line parallel to the real axis with imaginary part equal to $-\ii m$.
Furthermore, as mentioned above, Eq. (\ref{stattwo2yneg}) shows that at the stationary point $\ii y$ is real, and, as for the
integration in $x$, the saddle point application requires that the integral in $y$ be performed on a line parallel to the real axis.
The third and final step for the canonical case is:
\begin{eqnarray}
\label{canentrtwo_bneg}
&& \widetilde{s}_{{\rm can}}(\epsilon,m) = \min_{{\beta \le 0}} \Big\{ \beta \epsilon -\frac{\beta J}{2}m^2 + \frac{\beta K}{2} y^2 \nonumber \\
&&- \hat{\psi}(\beta,\beta J m + \varphi(\beta,y(\beta,m),m))
 - m \varphi(\beta,y(\beta,m),m) \Big\} \, .
\end{eqnarray}
The stationarity and stability conditions are given by
\begin{eqnarray}
\label{stattwo3neg}
&&\epsilon - \frac{J}{2}m^2 + \frac{K}{2}y^2 - \hat{\psi}_u - J m\hat{\psi}_v - \ii Ky\hat{\psi}_w = 0 \\
&&\hat{\psi}_{uu}{\hat{\psi}_{vv}} - \hat{\psi}_{uv}^2
+ 2 \ii Ky \left( \hat{\psi}_{uw}{\hat{\psi}_{vv}} - \hat{\psi}_{uv}\hat{\psi}_{vw} \right)
- \left(Ky\right)^2 \left( {\hat{\psi}_{vv}}\hat{\psi}_{ww} - \hat{\psi}_{vw}^2 \right) \nonumber \\
&-&\beta K \hat{\psi}_{vv} \frac{\left[ \hat{\psi}_{uw} - \frac{\hat{\psi}_{uv}\hat{\psi}_{vw}}{\hat{\psi}_{vv}} + \ii Ky
\left( \hat{\psi}_{ww} - \frac{\hat{\psi}_{vw}^2}{\hat{\psi}_{vv}} \right) \right]^2}
{1 + \beta K \left( \hat{\psi}_{ww} - \frac{\hat{\psi}_{vw}^2}{\hat{\psi}_{vv}} \right)} > 0 \, .
\label{stabtwo3neg}
\end{eqnarray}

We write for convenience, as before, the stability conditions of the problem (\ref{canentrtwoneg}). They are:
\begin{eqnarray}
\label{canstab1neg}
&&\hat{\psi}_{vv} < 0 \\
\label{canstab2neg}
&&1 + \beta K \left( \hat{\psi}_{ww} - \frac{\hat{\psi}_{vw}^2}{\hat{\psi}_{vv}} \right) > 0 \\
&&\hat{\psi}_{uu}{\hat{\psi}_{vv}} - \hat{\psi}_{uv}^2
+ 2 \ii Ky \left( \hat{\psi}_{uw}{\hat{\psi}_{vv}} - \hat{\psi}_{uv}\hat{\psi}_{vw} \right)
- \left(Ky\right)^2 \left( {\hat{\psi}_{vv}}\hat{\psi}_{ww} - \hat{\psi}_{vw}^2 \right)  \nonumber \\
&&-\beta K \hat{\psi}_{vv} \frac{\left[ \hat{\psi}_{uw} - \frac{\hat{\psi}_{uv}\hat{\psi}_{vw}}{\hat{\psi}_{vv}} + \ii Ky
\left( \hat{\psi}_{ww} - \frac{\hat{\psi}_{vw}^2}{\hat{\psi}_{vv}} \right) \right]^2}
{1 + \beta K \left( \hat{\psi}_{ww} - \frac{\hat{\psi}_{vw}^2}{\hat{\psi}_{vv}} \right)} > 0 \, .
\label{canstab3neg}
\end{eqnarray}

Going now to the microcanonical entropy, the second step for the problem (\ref{micrentrtwoneg}) is:
\begin{eqnarray}
\label{micrentrtwo_aneg}
&& \widetilde{s}_{{\rm micr}}(\epsilon,m) = \max_{x,y} \Big\{ \min_{{\beta \le 0}} \Big[
\beta \epsilon +\frac{\beta J}{2}x^2 + \frac{\beta K}{2} y^2 \nonumber \\
&&- \hat{\psi}(\beta,\ii \beta J x + \varphi(\beta,x,y,m),\ii \beta K y) - m \varphi(\beta,x,y,m) \Big] \Big\}  \, .
\end{eqnarray}
Minimization with respect to $\beta$ leads to the following stationarity and stability conditions:
\begin{eqnarray}
\label{stattwo4neg}
&&\epsilon + \frac{J}{2}x^2 + \frac{K}{2}y^2 - \hat{\psi}_u - \ii Jx\hat{\psi}_v - \ii Ky\hat{\psi}_w = 0 \\
%\label{stabtwo4neg}
&&\hat{\psi}_{uu}{\hat{\psi}_{vv}} - \hat{\psi}_{uv}^2
+ 2 \ii Ky \left( \hat{\psi}_{uw}{\hat{\psi}_{vv}} - \hat{\psi}_{uv}\hat{\psi}_{vw} \right)
- \left(Ky\right)^2 \left( {\hat{\psi}_{vv}}\hat{\psi}_{ww} - \hat{\psi}_{vw}^2 \right) > 0 \, . \nonumber \\
\label{stabtwo4neg}
&&
\end{eqnarray}
As we know, the stationarity condition is the same of the other extremal problem, and it gives $\beta$ as a function of
$(\epsilon,x,y,m)$. On the other hand, as for $\beta > 0$ the stability condition is different. From the stationarity condition
we obtain, in particular:
\begin{equation}
\label{betatwoyneg}
\beta_y = -\ii \beta K \frac{\hat{\psi}_{uw} - \frac{\hat{\psi}_{uv}\hat{\psi}_{vw}}{\hat{\psi}_{vv}} + \ii Ky
\left( \hat{\psi}_{ww} - \frac{\hat{\psi}_{vw}^2}{\hat{\psi}_{vv}}\right)}
{\hat{\psi}_{uu} - \frac{\hat{\psi}_{uv}^2}{\hat{\psi}_{vv}}
+ 2\ii Ky \left( \hat{\psi}_{uw} - \frac{\hat{\psi}_{uv}\hat{\psi}_{vw}}{\hat{\psi}_{vv}} \right)
- \left(Ky\right)^2 \left( \hat{\psi}_{ww} - \frac{\hat{\psi}_{vw}^2}{\hat{\psi}_{vv}} \right)} \, .
\end{equation}
We can now write the final step, given by:
\begin{eqnarray}
\label{micrentrtwo_bneg}
&& \widetilde{s}_{{\rm micr}}(\epsilon,m) =  \max_{x,y} \Big\{
\beta(\epsilon,x,y,m) \epsilon +\frac{\beta(\epsilon,x,y,m) J}{2}x^2 + \frac{\beta(\epsilon,x,y,m) K}{2} y^2 \nonumber \\
&&- \hat{\psi}(\beta(\epsilon,x,y,m),\ii \beta(\epsilon,x,y,m) J x + \varphi(\beta(\epsilon,x,y,m),x,y,m),\ii \beta(\epsilon,x,y,m) K y)
\nonumber \\
&&- m \varphi(\beta(\epsilon,x,y,m),x,y,m) \Big\}  \, .
\end{eqnarray}
The stationarity conditions are:
\begin{eqnarray}
\label{stattwo5xneg}
-\ii \hat{\psi}_v + x &=& 0 \\
\label{stattwo5yneg}
-\ii \hat{\psi}_w + y &=& 0 \, ,
\end{eqnarray}
equal respectively to (\ref{stattwo2xneg}) and (\ref{stattwo2yneg}), as expected. But the stability condition is not equal
to (\ref{stabtwo2neg}), being instead given by:
\begin{eqnarray}
\label{stabtwo5bneg}
&&1 + \beta K \left[ \hat{\psi}_{ww} - \frac{\hat{\psi}_{vw}^2}{\hat{\psi}_{vv}} \right] \\
&-&\beta K \frac{\left[ \hat{\psi}_{uw} - \frac{\hat{\psi}_{uv}\hat{\psi}_{vw}}{\hat{\psi}_{vv}} + \ii Ky
\left( \hat{\psi}_{ww} - \frac{\hat{\psi}_{vw}^2}{\hat{\psi}_{vv}}\right) \right]^2}
{\hat{\psi}_{uu} - \frac{\hat{\psi}_{uv}^2}{\hat{\psi}_{vv}}
+ 2\ii Ky \left( \hat{\psi}_{uw} - \frac{\hat{\psi}_{uv}\hat{\psi}_{vw}}{\hat{\psi}_{vv}} \right)
- \left(Ky\right)^2 \left( \hat{\psi}_{ww} - \frac{\hat{\psi}_{vw}^2}{\hat{\psi}_{vv}} \right)} > 0 \, . \nonumber
\end{eqnarray}

The summary of the stability conditions of the problem (\ref{micrentrtwoneg}) is:
\begin{eqnarray}
\label{micrstab1neg}
&&\hat{\psi}_{vv} < 0 \\
\label{micrstab2neg}
&&\hat{\psi}_{uu}{\hat{\psi}_{vv}} - \hat{\psi}_{uv}^2
+ 2\ii Ky \left( \hat{\psi}_{uw}{\hat{\psi}_{vv}} - \hat{\psi}_{uv}\hat{\psi}_{vw} \right) \nonumber \\
&&- \left(Ky\right)^2 \left( {\hat{\psi}_{vv}}\hat{\psi}_{ww} - \hat{\psi}_{vw}^2 \right) > 0 \\
\label{micrstab3neg}
&&1 + \beta K \left[ \hat{\psi}_{ww} - \frac{\hat{\psi}_{vw}^2}{\hat{\psi}_{vv}} \right]  \\
&&-\beta K \frac{\left[ \hat{\psi}_{uw} - \frac{\hat{\psi}_{uv}\hat{\psi}_{vw}}{\hat{\psi}_{vv}} + \ii Ky
\left( \hat{\psi}_{ww} - \frac{\hat{\psi}_{vw}^2}{\hat{\psi}_{vv}}\right) \right]^2}
{\hat{\psi}_{uu} - \frac{\hat{\psi}_{uv}^2}{\hat{\psi}_{vv}}
+ 2\ii Ky \left( \hat{\psi}_{uw} - \frac{\hat{\psi}_{uv}\hat{\psi}_{vw}}{\hat{\psi}_{vv}} \right)
+ \left(Ky\right)^2 \left( \hat{\psi}_{ww} - \frac{\hat{\psi}_{vw}^2}{\hat{\psi}_{vv}} \right)}  > 0 \, . \nonumber
\end{eqnarray}

\subsection{Unified treatment of temperatures of both signs}
\label{secbothtemp}

From the above computations we see that it is possible to treat at the same time both positive and negative temperatures. It is
sufficient to put from the start $x=m$ for $\beta>0$ and $\ii x = m$ for $\beta <0$, and identifying $y$ for $\beta >0$ with
$\ii y$ for $\beta < 0$. Since in this way it is not necessary to optimize with respect to $x$, the numerical procedure would
be shorter. Thus, the extremal problems to study would be:
\begin{equation}
\widetilde{\phi}(\beta,m) = \min_y \left[ \max_{\varphi} \left( \frac{\beta J}{2}m^2 + \frac{\beta K}{2} y^2
+ \hat{\psi}(\beta,\beta J m +\varphi,\beta K y) + m \varphi \right) \right] \, ,
\label{exphitwoboth}
\end{equation}
for the rescaled free energy,
\begin{eqnarray}
\label{canentrtwoboth}
&&\,\,\,\,\,\,\,\,\,\,\,\,\,\,\,\,\,\,\,\,\,\,\,\,\,\,\,\,\,\,\, \widetilde{s}_{{\rm can}}(\epsilon,m) = \\
&=& \min_{\beta} \left\{ \max_y \left[ \min_{\varphi} \left(
\beta \epsilon -\frac{\beta J}{2}m^2 - \frac{\beta K}{2} y^2
- \hat{\psi}(\beta,\beta J m + \varphi,\beta K y) - m \varphi \right) \right] \right \} \, , \nonumber
\end{eqnarray}
for the canonical entropy, and
\begin{eqnarray}
\label{micrentrtwoboth}
&&\,\,\,\,\,\,\,\,\,\,\,\,\,\,\,\,\,\,\,\,\,\,\,\,\,\,\,\,\,\,\, \widetilde{s}_{{\rm micr}}(\epsilon,m) = \\
&=& \max_y \left\{ \min_{\beta} \left[ \min_{\varphi} \left(
\beta \epsilon -\frac{\beta J}{2}m^2 - \frac{\beta K}{2} y^2
- \hat{\psi}(\beta,\beta J m + \varphi,\beta K y) - m \varphi \right) \right] \right \} \, , \nonumber
\end{eqnarray}
for the microcanonical entropy. The search for the extremal points would proceed as before, with the stationary conditions summarized by:
\begin{eqnarray}
\label{cond1both}
\hat{\psi}_v + m &=& 0 \\
\label{cond3both}
\hat{\psi}_w + y &=& 0 \\
\label{cond4both}
\epsilon -\frac{J}{2}m^2 -\hat{\psi}_u - J m\hat{\psi}_v - K y\hat{\psi}_w &=& 0 \, ,
\end{eqnarray}
while the stability conditions would be
\begin{eqnarray}
\label{canstab1both}
&&\hat{\psi}_{vv} < 0 \\
\label{canstab2both}
&&1 + \beta K \left( \hat{\psi}_{ww} - \frac{\hat{\psi}_{vw}^2}{\hat{\psi}_{vv}} \right) > 0 \\
&&\hat{\psi}_{uu}{\hat{\psi}_{vv}} - \hat{\psi}_{uv}^2
+ 2 Ky \left( \hat{\psi}_{uw}{\hat{\psi}_{vv}} - \hat{\psi}_{uv}\hat{\psi}_{vw} \right)
+ \left(Ky\right)^2 \left( {\hat{\psi}_{vv}}\hat{\psi}_{ww} - \hat{\psi}_{vw}^2 \right)  \nonumber \\
&&-\beta K \hat{\psi}_{vv} \frac{\left[ \hat{\psi}_{uw} - \frac{\hat{\psi}_{uv}\hat{\psi}_{vw}}{\hat{\psi}_{vv}} + Ky
\left( \hat{\psi}_{ww} - \frac{\hat{\psi}_{vw}^2}{\hat{\psi}_{vv}} \right) \right]^2}
{1 + \beta K \left( \hat{\psi}_{ww} - \frac{\hat{\psi}_{vw}^2}{\hat{\psi}_{vv}} \right)} > 0 \, ,
\label{canstab3both}
\end{eqnarray}
for the canonical problem (\ref{canentrtwoboth}), and
\begin{eqnarray}
\label{micrstab1both}
&&\hat{\psi}_{vv} < 0 \\
\label{micrstab2both}
&&\hat{\psi}_{uu}{\hat{\psi}_{vv}} - \hat{\psi}_{uv}^2
+ 2 Ky \left( \hat{\psi}_{uw}{\hat{\psi}_{vv}} - \hat{\psi}_{uv}\hat{\psi}_{vw} \right) \nonumber \\
&&+ \left(Ky\right)^2 \left( {\hat{\psi}_{vv}}\hat{\psi}_{ww} - \hat{\psi}_{vw}^2 \right) > 0 \\
\label{micrstab3both}
&&1 + \beta K \left[ \hat{\psi}_{ww} - \frac{\hat{\psi}_{vw}^2}{\hat{\psi}_{vv}} \right]  \\
&&-\beta K \frac{\left[ \hat{\psi}_{uw} - \frac{\hat{\psi}_{uv}\hat{\psi}_{vw}}{\hat{\psi}_{vv}} + Ky
\left( \hat{\psi}_{ww} - \frac{\hat{\psi}_{vw}^2}{\hat{\psi}_{vv}}\right) \right]^2}
{\hat{\psi}_{uu} - \frac{\hat{\psi}_{uv}^2}{\hat{\psi}_{vv}}
+ 2 Ky \left( \hat{\psi}_{uw} - \frac{\hat{\psi}_{uv}\hat{\psi}_{vw}}{\hat{\psi}_{vv}} \right)
+ \left(Ky\right)^2 \left( \hat{\psi}_{ww} - \frac{\hat{\psi}_{vw}^2}{\hat{\psi}_{vv}} \right)}  > 0 \, , \nonumber
\end{eqnarray}
for the microcanonical problem (\ref{micrentrtwoboth}). However, now the search could be extended also to $\beta <0$.

At this point we can discuss the issue of ensemble inequivalence, on the basis of the comparison between the stability conditions of
the two problems. We start by noting what can be deduced from the expression of $\hat{\psi}(\beta,\beta J x + \varphi,\beta K y)$ given
in Eq. (\ref{psihatdeftwo}). In fact, $\hat{\psi}(\beta,\beta J x + \varphi,\beta K y)$ is proportional to minus the logarithm of the
right hand side of that equation, and then it is easy to see that for any $x$, $y$ and $\varphi$ the total second derivative of
$\hat{\psi}$ with respect to $\beta$ is negative. From this, in turn, one obtains that the left hand side of Eq. (\ref{micrstab2both})
is always positive, i.e., it is a stability condition of the microcanonical problem that is always satisfied. The consequences are the
following, taking into account that for both problems the first stability condition is $\hat{\psi}_{vv}<0$.
Let us first suppose to have a solution of the canonical problem, i.e., to have a stationary point where the three stability conditions
(\ref{canstab1both})-(\ref{canstab3both}) are satisfied\footnote{It is not difficult to see that for $\beta > 0$ the conditions
(\ref{canstab1both}) and (\ref{canstab2both}), taken together, imply the condition (\ref{canstab3both}); this is not true for
$\beta < 0$.}. Then also the microcanonical stability condition (\ref{micrstab3both}) is satisfied. This can be seen, e.g., in the
following way: the left hand side of (\ref{micrstab3both}) is equal to the left hand side of (\ref{canstab3both}) multiplied by the
ratio of the left hand side of (\ref{canstab2both}) and the left hand side of (\ref{micrstab2both}); since this ratio is positive, this
implies that (\ref{micrstab3both}) is satisfied. In other words, any stable canonical equilibrium state is also a stable microcanonical
equilibrium state. This is consistent with a general result obtained, e.g., through large deviation techniques \cite{PR}. On the other
hand, it is possible to satisfy the stability condition (\ref{micrstab3both}) of the microcanonical problem without satisfying the
stability condition (\ref{canstab2both}) of the canonical problem (and then, for what we have seen, neither (\ref{canstab3both}) would
be satisfied); namely, there can be stable microcanonical equilibrium states that are not stable canonical equilibrium states. Nevertheless,
although for the values of $(\epsilon,m)$ where the latter situation is verified there are not stable canonical equilibrium states, still
the extremal problem (\ref{canentrtwoboth}) can be satisfied. This occurs since in that case the function $y(\beta)$ defined by the
stationarity condition (\ref{cond3both}) has a point of discontinuity in its derivative with respect to $\beta$, given by
Eq. (\ref{ytwobeta}), and for a range of $\epsilon$ values the minimization problem (\ref{canentrtwoboth}) is satisfied by the
$\beta$ value of the point of discontinuity. This is associated with the occurrence of a first order phase transition in the canonical
ensemble: there are not stable canonical equilibrium states for the $\epsilon$ values in that range (but only for the two values at the
extremes of the range), and the computed canonical entropy $\widetilde{s}_{{\rm can}}(\epsilon,m)$ has a straight line segment for
that $\epsilon$ range. In that range we have strictly
$\widetilde{s}_{{\rm micr}}(\epsilon,m) < \widetilde{s}_{{\rm can}}(\epsilon,m)$\footnote{In a system with only short-range interactions,
where ensembles are equivalent and the two entropies are always equal, $\widetilde{s}_{{\rm micr}}(\epsilon,m)$ would have the same
straight line segment, and for the values of $\epsilon$ inside the range of the segment the equilibrium states, in both ensembles, would
be realized with a phase separation, something that does not occur in long-range systems.}.

In Appendix A we provide an alternative derivation of the expression of the maximization problem (\ref{micrentrtwoboth}) for the
microcanonical entropy, using a procedure based on large deviation techniques.

\subsection{The limit $J\to 0$ and $K \to 0$}
\label{secjto0}

We want to show that in the limit $J\to 0$ and $K \to 0$ our expressions go continuously to those one would expect for the
Hamiltonian (\ref{modhamext}) in absence of mean-field terms, i.e., when the system becomes a short-range one. We first note that the
Hubbard-Stratonovich equality (\ref{hubtran}) is valid also in the limit $a \to 0$, as it can be easily checked. This implies that
for $J \to 0$ and $K \to 0$ the expressions (\ref{canpart0two}) and (\ref{micrpart0two}) remain valid by simply removing, on the right
hand side, the integrations over $x$ and $y$ with the corresponding prefactors, and by putting $J=0$ and $K=0$ in the integrand. In
particular, now the function $\hat{\psi}$ would appear as $\hat{\psi}(\beta,\varphi,0)$, with
$\exp \left[ -N \hat{\psi}(\beta,\varphi,0) \right]$ being equal to the partition function of the short-range system subject to a
magnetic field $h$ with $\varphi$ playing the role of $\beta h$. Correspondingly, in the expressions (\ref{exphitwo}),
(\ref{canentrtwo}) and (\ref{micrentrtwo}) for the rescaled free energy and the entropy, the variables $x$ and $y$ would disappear,
and we would get:
\begin{equation}
\widetilde{\phi}(\beta,m) = \max_{\varphi} \left( \hat{\psi}(\beta,\varphi,0) + m \varphi \right)
\label{exphij0}
\end{equation}
and
\begin{equation}
\widetilde{s}(\epsilon,m) = \min_{\beta} \left[ \min_{\varphi} \left(
\beta \epsilon - \hat{\psi}(\beta,\varphi,0) - m \varphi \right) \right] \, .
\label{canentrj0}
\end{equation}
The latter equation holds for both the canonical and microcanonical cases.

Then, we can conclude this subsection by observing that the method presented in this paper works then both for short-range interactions
(when both $J$ and $K$ are equal to $0$) and for long-range interactions. For the former the problem of determining the entropy at
fixed magnetization is solved once that the partition function in a magnetic field is known, as expected. However, in short-range
models there is not the issue whether the microcanonical entropy and the canonical one are different: they are the same in the
thermodynamical limit \cite{Ruelle}. Similarly, one can see that microcanonical and canonical entropies at fixed magnetization are
also the same in short-range models. For this reason, we devote the section of the implementation to models with $J \neq 0$, where ensemble
inequivalence can occur and it is interesting to compute and compare canonical and microcanonical entropies at fixed magnetization.
As anticipated, in the implementation we consider systems with only one mean-field term, thus with $K=0$.

\subsection{The model with only one mean-field term}
\label{model_one}

In the next section we will implement the method in a model where the Hamiltonian contains only one mean-field term, the one
proportional to $-N\hat{m}^2$. The corresponding expressions would be obtained by the ones derived above by putting $K=0$ and in which
there would be no integration over the $y$ auxiliary variable and the corresponding optimization with respect to it.

However, we write in the following the optimizations problems and the corresponding stationarity and stability conditions, since an
interesting issue arises. From Eq. (\ref{exphitwoboth}) with $K=0$ and with no optimization with respect to $y$, we obtain
\begin{equation}
\widetilde{\phi}(\beta,m) = \max_{\varphi} \left( \frac{\beta J}{2}m^2 + \hat{\psi}(\beta,\beta J m +\varphi,0)
+ m \varphi \right) \, .
\label{exphi_one}
\end{equation}
In the same way, for the entropies, from Eq. (\ref{canentrtwoboth}) and (\ref{micrentrtwoboth}) we obtain in both cases
\begin{equation}
\widetilde{s}(\epsilon,m) = \min_{\beta} \left[ \min_{\varphi} \left(
\beta \epsilon -\frac{\beta J}{2}m^2 - \hat{\psi}(\beta,\beta J m + \varphi,0) - m \varphi \right) \right] \, ,
\label{micrentr_one}
\end{equation}
expression which is then valid for both ensembles. The stationarity and stability conditions of the latter problem are
\begin{eqnarray}
\label{statstab4}
\hat{\psi}_v + m = 0 &\,\,\,\,\,\,\,\,\,\,\,\,\,& \hat{\psi}_{vv} < 0 \\
\label{statstab5}
\epsilon - \frac{J}{2}m^2 - \hat{\psi}_u - J m\hat{\psi}_v = 0
&\,\,\,\,\,\,\,\,\,\,\,\,\,& \hat{\psi}_{uu}\hat{\psi}_{vv} - \hat{\psi}^2_{uv} > 0 \, .
\end{eqnarray}

The fact that now we have the same optimization problem for both ensembles, allows to make the following interesting observation.
For the class of systems where the long-range interaction is given only by a term proportional to $-N\hat{m}^2$, the entropy at
fixed magnetization is the same for the canonical and the microcanonical case. From the equality
$\widetilde{s}_{{\rm micr}}(\epsilon,m) = \widetilde{s}_{{\rm can}}(\epsilon,m)$,
one might superficially conclude that there is ensemble equivalence. However, equivalence occurs when we have
$s_{{\rm micr}}(\epsilon) = s_{{\rm can}}(\epsilon)$, and it is not automatically guaranteed that this latter equality is
verified when the former equality, $\widetilde{s}_{{\rm micr}}(\epsilon,m) = \widetilde{s}_{{\rm can}}(\epsilon,m)$, holds.
In a moment we provide a concrete example of this fact, but first we describe the mathematical reason that can explain why
it is possible to have $s_{{\rm micr}}(\epsilon) < s_{{\rm can}}(\epsilon)$ for one or more ranges of the energy\footnote{We
remind that, because of the properties of min-max extremal problems \cite{PR}, in general one has
$s_{{\rm micr}}(\epsilon) \le s_{{\rm can}}(\epsilon)$.} even when
$\widetilde{s}_{{\rm micr}}(\epsilon,m) = \widetilde{s}_{{\rm can}}(\epsilon,m)$. The microcanonical entropy
$s_{{\rm micr}}(\epsilon)$ is obtained by maximizing $\widetilde{s}_{{\rm micr}}(\epsilon,m)$ with respect to $m$, i.e.,
$s_{{\rm micr}}(\epsilon) = \max_m \left[ \widetilde{s}_{{\rm micr}}(\epsilon,m) \right]$; however, the canonical entropy
$s_{{\rm can}}(\epsilon)$ is not obtained performing the analogous maximization of
$\widetilde{s}_{{\rm can}}(\epsilon,m)$. In fact, $s_{{\rm can}}(\epsilon)$ is given by the general thermodynamic relation
$s_{{\rm can}}(\epsilon) = \min_{\beta} \left[ \beta \epsilon - \phi (\beta) \right]$, where, in our case,
$\phi(\beta) = \min_m \widetilde{\phi}(\beta,m)$, with the latter function obtained in turn from
$\widetilde{s}_{{\rm micr}}(\epsilon,m)$ through the minimization problem in Eq. (\ref{lftransf}). These different extremization
procedures to obtain $s_{{\rm micr}}(\epsilon)$ and $s_{{\rm can}}(\epsilon)$  can thus lead to different functions. An
example can be found in the Blume-Capel model. It is a simplified version of the model cited in footnote \ref{footbeg}, and in
which the general Hamiltonian (\ref{modhamext}) has $K=0$ (i.e., the only mean-field term is the one proportional to
$-N\hat{m}^2$) and the function $U([S_i])$ simply given by $\Delta S_i^2$ ($\Delta$ is a positive parameter); the spins take
the values $-1,0,1$. For this simple model direct counting can easily be performed, and in Ref. \cite{BMR2001} it is
shown that ensemble inequivalence occurs, since there are ranges of the energy where
$s_{{\rm micr}}(\epsilon) < s_{{\rm can}}(\epsilon)$. However, it is also easy to see explicitly that
$\widetilde{s}_{{\rm micr}}(\epsilon,m) = \widetilde{s}_{{\rm can}}(\epsilon,m)$. In fact, a direct computation allows to
obtain
\begin{eqnarray}
\widetilde{s}_{{\rm micr}}(\epsilon,m) &=& -\left[ 1- b(\epsilon,m)\right] \ln \left[ 1-b(\epsilon,m)\right]
-\frac{1}{2}\left[b(\epsilon,m)+m\right]\ln \left[b(\epsilon,m)+m\right] \nonumber \\
&&-\frac{1}{2}\left[b(\epsilon,m)-m\right]\ln \left[b(\epsilon,m)-m\right] + b(\epsilon,m)\ln 2 \, ,
\label{blcapentr}
\end{eqnarray}
where $b(\epsilon,m) \equiv \frac{\epsilon}{\Delta} + \frac{m^2}{2\Delta}$. Obviously, $\epsilon$ and $m$ can vary
in ranges for which the argument of all the logarithms in (\ref{blcapentr}) are non-negative. For any allowed value
of $\epsilon$ and $m$, this function is concave in $\epsilon$. Therefore the Legendre-Fenchel transform (\ref{lftransf})
is invertible, assuring that $\widetilde{s}_{{\rm micr}}(\epsilon,m) = \widetilde{s}_{{\rm can}}(\epsilon,m)$,
in agreement with the general result of this section, i.e., that in models with only the mean-field term proportional
to $-N\hat{m}^2$ we have, for both ensembles, the same entropy
$\widetilde{s}(\epsilon,m)$ given in Eq. (\ref{micrentr_one}).

In conclusion, ensemble inequivalence can occur even when $\widetilde{s}_{{\rm micr}}(\epsilon,m)$ and
$\widetilde{s}_{{\rm can}}(\epsilon,m)$ are equal; even when $\widetilde{s}_{{\rm micr}}(\epsilon,m)$ is concave in
$\epsilon$ for any $m$, so that the Legendre-Fenchel transform (\ref{lftransf}) is invertible, $s_{{\rm micr}}(\epsilon)$
can be non-concave, so that the Legendre-Fenchel transform (\ref{lftransfgen}) is not invertible.

\section{Implementation of the method}
\label{secmodels}

As anticipated above, we implement here our computational method to an Ising spin model ($S_i = \pm 1$) on a one-dimensional
lattice, described by a Hamiltonian of the type (\ref{modhamext}) in which the short-range term is given by
$U([S_i])=-(K_1/2) S_i S_{i+1}-(K_2/2) S_i S_{i+2}$. Furthermore, the coefficient of the quadrupole term, $K$, is set equal to
zero (in any case, the quadrupole term for Ising spins would give a constant trivial contribution). Therefore, we are going to
treat a system with only the mean-field term proportional to $-N\hat{m}^2$. For convenience, we write the explicit form of
the Hamiltonian, i.e.:
\begin{equation}
\label{HAM12}
H=-\frac{J}{2N}\left(\sum_{i=1}^N S_i\right)^2 - \frac{K_1}{2} \sum_{i=1}^N S_i S_{i+1} -
\frac{K_2}{2}\sum_{i=1}^N S_i S_{i+2} \, ,
\end{equation}
where periodic boundary conditions are assumed. Thus, the short-range part of this Hamiltonian has a nearest-neighbour interaction
term (with coefficient $K_1$) and next-nearest-neighbour interaction term (with coefficient $K_2$). It is the simplest Hamiltonian
having a long-range interaction term and a short-range part with an internal structure, with the $K_1$ and $K_2$ parts of the model
possibly competing.

To show the intricacies of the calculation of microcanonical entropy at fixed magnetization using direct counting, also in relatively
simple models like this one, and the convenience of the implementation of the method presented in this work, we adopt the following
strategy. Using the results given in Ref. \cite{Gori11}, we will first describe, in subsection 4.1, what would be the procedure to follow
to obtain a direct counting evaluation of $\widetilde{s}_{{\rm micr}}(\epsilon,m)$. From this description and the expressions to use in such
a procedure (the interested reader can find in Appendix B a summary of the results of Ref. \cite{Gori11} and of the relevant expressions),
it will be evident that the actual numerical computations are quite cumbersome, so that the optimization method introduced
in section 3 has to be preferred. Then, in subsection 4.2 we will show the evaluation of $\widetilde{s}_{{\rm micr}}(\epsilon,m)$
with the method of this work. This will be done first for the case with $K_2=0$; in this case, the direct counting is simple,
and it was done in Ref. \cite{Mukamel05}, a work that was primarily devoted to the study of ensemble inequivalence. Thus, we will also show
the comparison of our computation with that obtained with direct counting. Afterwords, we will show our results for some selected
values of $K_2 \neq 0$. In this case the direct counting becomes rather involved, and has not been done up to now. In order to
provide a comparison, we have performed the direct counting evaluation; this is shown in Appendix C.

\subsection{The procedure for the direct counting evaluation}
\label{proc_direct}

Ref. \cite{Gori11} gives a general procedure for a computation of the microcanonical entropy of translationally invariant
one-dimensional Ising spin systems with short-range interactions. Restricting to models with only two-spin interactions, a short-range
Hamiltonian should have the form:
\begin{equation}
H_{SR}=-\frac{1}{2}\sum_{r=1}^R K_r \sum_i S_i S_{i+r} \, .
\label{HSR}
\end{equation}
The subscript $(SR)$ denotes that the Hamiltonian describes a short-range system. The procedure can be extended to models where also
a long-range interaction, like the first term in the Hamiltonian (\ref{HAM12}), is present.  For convenience we will denote this
long-range part of the Hamiltonian with $H_{LR}$. Therefore in (\ref{HAM12}) we have $H = H_{LR} + H_{SR}$, where in this case
$H_{SR}$ is a particular form of (\ref{HSR}) having $R=2$. Despite
the simplicity of the model (\ref{HAM12}), it displays a remarkably rich phase diagrams in the canonical ensemble \cite{Campa19}.

While more details for the case with generic $R$ are given in Appendix B, let us here consider the direct counting procedure for the
Hamiltonian (\ref{HAM12}). We begin with the simple case in which $K_2=0$.

\subsubsection{$K_2=0$}
\label{K2_0}
It is possible to compute the number of spin configurations that have given values $m$ and $g_1$ of, respectively, the magnetization
$\hat{m}$ and the nearest-neighbour correlation function $\hat{g}_1$:
\begin{align}
& \hat{m}=\frac{1}{N}\sum_{i=1}^N S_i \\
& \hat{g}_1=\frac{1}{N}\sum_{i=1}^N S_i S_{i+1} \, .
\end{align}
The logarithm, divided by $N$, of such number of spin configurations can be denoted with $s(m,g_1)$, and called the entropy
as a function of $m$ and $g_1$. In the thermodynamic limit it is given by \cite{Gori11}:
\begin{align}\label{entropy1}
  & s(m,g_1) =
  \nonumber \\
& - \frac{1+2 m+g_1}{4} \ln\frac{1+2 m+g_1}{4} - \frac{1-2 m+g_1}{4}\ln\frac{1-2 m+g_1}{4}
  \nonumber \\
& - \frac{1-g_1}{2} \ln\frac{1-g_1}{4} + \frac{1+m}{2} \ln\frac{1+m}{2} + \frac{1-m}{2} \ln\frac{1-m}{2}\,.
\end{align}
It follows that $s(m,g_1)$ is defined within the convex polytope (polygon) defined by the following constraints:
\begin{align}
1 + 2 m + g_1 > 0, \qquad 1 - 2 m + g_1 > 0, \qquad 1 - g_1 > 0 \, .
\end{align}
As noted above, the expression for $s(m,g_1)$ was obtained already in Ref. \cite{Mukamel05}. It must be emphasized that,
although $s(m,g_1)$ is naturally called an entropy, being given by the logarithm of a number of configurations,
it is not the usual thermodynamic entropy, that has to depend on the energy $\epsilon$, like the function of our interest,
$\widetilde{s}_{\rm micr}(\epsilon,m)$. To obtain the latter from $s(m,g_1)$, one has to express the energy as a function of
$m$ and of the correlation $g_1$, as we now explain (obviously the same remark is valid also for the more complex case
with $K_2 \neq 0$, discussed below).

From the Hamiltonian (\ref{HAM12}) with $K_2=0$ one obatins the energy per spin at fixed values $m$ and $g_1$. It is given by:
\begin{equation}
 e(m,g_1) = -\frac{J}{2}m^2 - \frac{K_1}{2} g_1 \,.
 \label{en1}
\end{equation}
To find $\widetilde{s}_{\rm micr}(\epsilon,m)$ one has to fix $e(m,g_1)\equiv \epsilon$, express from (\ref{en1}) $g_1$ as a function
of $\epsilon$ and $m$, and then substitute $g_1(\epsilon,m)$ in Eq. (\ref{entropy1}). This can be done producing results in agreement
with those presented in \cite{Mukamel05}.

We now go to the model with $K_2 \neq 0$, i.e., the complete Hamiltonian (\ref{HAM12}); the direct counting procedure becomes much
more involved.

\subsubsection{$K_2 \neq 0$}\label{J_K1_K2_model}
\label{morecompmod}

Using the procedure described in Appendix B, one finds the number of spin configurations having given values $m$, $g_1$, $g_2$
and $t$ of, respectively, the already defined magnetization $\hat{m}$ and nearest-neighbour correlation function $\hat{g}_1$, and 
of the other correlation functions $\hat{g}_2$ and $\hat{t}$; they are given by: 
\begin{align}
  & \hat{m}=\frac{1}{N}\sum_{i=1}^N S_i  & \hat{g}_1=\frac{1}{N}\sum_{i=1}^N S_i S_{i+1} \\
  & \hat{g}_2=\frac{1}{N}\sum_{i=1}^N S_i S_{i+2} & \hat{t}=\frac{1}{N}\sum_{i=1}^N S_i S_{i+1} S_{i+2} \, .
\end{align}
The fixed value of one of this quantity can conveniently be denoted with the average of the corresponding spin correlation, e.g.,
$g_2 = \langle S_i S_{i+2} \rangle$, where translational invariance assures that the average actually does not depend on $i$.
The logarithm, divided by $N$, of the number of the configurations at fixed $(m,g_1,g_2,t)$, i.e., the entropy as a function
of these variables, in the thermodynamic limit is obtained as the following long expression:
{\scriptsize
  \begin{align}
    \label{entropy2}
& s(m,g_1,g_2,t) = \nonumber \\
& -\frac{1+m-g_2-t}{4} \ln\frac{1+m-g_2-t}{8}-\frac{1+m-2 g_1+g_2-t}{8} \ln\frac{1+m-2 g_1+g_2-t}{8} \nonumber \\
& -\frac{1-3 m+2 g_1+g_2-t}{8} \ln\frac{1-3 m+2 g_1+g_2-t}{8}-\frac{1-m-g_2+t}{4} \ln\frac{1-m-g_2+t}{8} \nonumber \\
& -\frac{1-m-2 g_1+g_2+t}{8} \ln\frac{1-m-2 g_1+g_2+t}{8}-\frac{1+3 m+2 g_1+g_2+t}{8} \ln\frac{1+3 m+2 g_1+g_2+t}{8} \nonumber \\
& + \frac{1+2 m+g_1}{4} \ln\frac{1+2 m+g_1}{4} + \frac{1-2 m+g_1}{4}\ln\frac{1-2 m+g_1}{4}
    + \frac{1-g_1}{2} \ln\frac{1-g_1}{4} \,.
\end{align}}
From this entropy, which is a function of the four quantities $(m,g_1,g_2,t)$, we can obtain the entropy as a function of $(m,g_1,g_2)$
by maximizing $s(m,g_1,g_2,t)$ with respect to $t$, i.e.
\begin{align}
& s(m,g_1,g_2) = s(m,g_1,g_2,t_0)\nonumber \\
& \left.\frac{\partial s(m,g_1,g_2,t)}{\partial t} = 0 \right|_{t=t_0} \,.
\label{derivata}
\end{align}
An explicit expression for this entropy is still obtainable, involving the
real solution of the third order equation in $t_0$:
\begin{align}
  & t_0^3-m(2g_1+g_2)t_0^2+[(1-2 g_1^2+2 g_1^2 g_2-g_2^2)+m^2(-3+4 g_1+2 g_2)] t_0
  \nonumber \\
  & + m[-2 g_1+2 g_1^2-g_2+4 g_1 g_2-2 g_1^2 g_2-2 g_1 g_2^2+g_2^3+m^2(2-2 g_1-g_2)] = 0 \, ,
  \label{cub}
\end{align}
maximizing $s(m,g_1,g_2,t)$.

One finds from (\ref{entropy2}) that $s(m,g_1,g_2)$ is defined within the convex polytope (polyhedron) defined by the following
constraints:
\begin{align}
& 1 + 2 m + g_1 > 0 \qquad 1 - 2 m + g_1 > 0 \qquad 1 - g_1 > 0 &\\
& 1 + 2 m + g_2 > 0 \qquad 1 - 2 m + g_2 > 0 \qquad 1 - g_2 > 0 & \\
  & 1 + 2 g_1 + g_2 > 0 \qquad 1 - 2 g_1 + g_2 > 0 \, . \qquad
  \label{region}
\end{align}
Notice that for certain ranges of the parameters $m,g_1,g_2$ there are three real roots of (\ref{cub}), but in these ranges direct
inspection shows that for only one of these three roots the entropy is defined -- i.e., all arguments of the logarithms in
(\ref{entropy2}) are positive.

The energy per spin at fixed values of $m$, $g_1$, $g_2$ and $t$ is given by:
\begin{equation}
  \label{en2}
 e(m,g_1,g_2,t) = -\frac{J}{2} m^2 - \frac{K_1}{2} g_1 - \frac{K_2}{2} g_2 \, ,
\end{equation}
that actually does not depend on $t$, since the Hamiltonian (\ref{HAM12}) has no contribution coming from three spin terms
$\propto S_i S_{i+1} S_{i+2}$ (the extension to this case is straightforward). At this point, to find
$\widetilde{s}_{\rm micr}(\epsilon,m)$ one has to fix $e\equiv \epsilon$ where $e=e(m,g_1,g_2,t=t_0)$. Then from (\ref{en2}) one
has to express, say, $g_2$ [or, if one wants so, $g_1$] as a function of $\epsilon$, $m$ and $g_1$ [or, respectively, as a function of
$\epsilon$, $m$ and $g_2$]. Then one has to substitute $g_2(\epsilon,m;g_1)$ in Eq. (\ref{entropy2}). The entropy will now depend on
$\epsilon,m$ and $g_1$: maximizing with respect to $g_1$ will give $\widetilde{s}_{\rm micr}(\epsilon,m)$.

It is then clear that, despite the expression of the entropy is known, already in this simple case one can (painfully) realize that the
computations from expression (\ref{entropy2}) is quite cumbersome. The analysis also clearly shows how difficult is to generalize it
to more complicated
forms of coupling, where the equations fixing the further correlation functions and the expression of the entropy become rapidly very
involved, e.g. when a next-to-next-nearest-neighbour coupling is added. For example, assuming an Hamiltonian of the form
$$H=-\frac{J}{2N}\sum_{i,j} S_i S_j - \frac{K_1}{2} \sum_{i} S_i S_{i+1} -
  \frac{K_2}{2} \sum_{i} S_i S_{i+2}-
  \frac{K_3}{2} \sum_{i} S_i S_{i+3}\,,$$
one has an energy $e=e(m,g_1,g_2,g_3)$ with $g_3=\langle S_i S_{i+3} \rangle$. The entropy $s=s(m,g_1,g_2,g_3,t,q_1,q_2,q_3)$ is a
function of the already introduced quantities $m$, $g_1$, $g_2$, $g_3$ and $t$, and of the other quantities
$q_1=\langle S_{i} S_{i+1} S_{i+3}\rangle$, $q_2=\langle S_{i} S_{i+2} S_{i+3}\rangle$, and
$q_3=\langle S_{i} S_{i+1} S_{i+2} S_{i+3}\rangle$, Despite having a quite convoluted expression, it can be determined using the method of
Appendix B. Now, to arrive at $\widetilde{s}_{\rm micr}(\epsilon,m)$, one has first to fix $t,q_1,q_2,q_3$ by maximizing
$s(m,g_1,g_2,g_3,t,q_1,q_2,q_3)$ with respect to these variables from the
conditions $\partial s/\partial t=\partial s/\partial q_1=\partial s/\partial q_2=\partial s/\partial q_3=0$. Then fixing the energy to a
certain value $\epsilon$, from $\epsilon=e(m,g_1,g_2,g_3)$ one gets, say, $g_3=g_3(m,g_1,g_2)$ and then an entropy $s$ as a function of
$(m,g_1,g_2)$. Maximizing with respect to $g_1$ and $g_2$ one finally obtains the microcanonical entropy at fixed magnetization,
$\widetilde{s}_{\rm micr}(\epsilon,m)$. So, one
sees that increasing the range of the couplings and/or adding multi-spin interactions, one has first to eliminate the couplings not
present in the Hamiltonian, and then -- after using the expression of the energy -- maximize over the remaining, which is of course a
complicated task. This can be seen as a ``direct counting'', since the explicit expression of the entropy is used; but one concludes that
the method presented in the section 3 is more practical, since it involves only the extremization with respect to a given number
of variables (depending on the number of auxiliary variables), independent from the range of the short-range interactions. What is
actually increasing when more couplings are included is the size of the transfer matrix, whose only the largest eigenvalue is needed. Of
course, the use of the transfer matrix would be present also in the computation of only canonical quantities. At variance, in the direct
counting one has to maximize with respect to a growing number of variables, as the example now discussed shows. We have however to observe
that the direct counting, as shown in Appendix B, gives not only the microcanonical entropy at fixed magnetization but also the
values -- at given energy and magnetization -- of all independent correlation functions in the unit cell, e.g., in the considered case
with $K_1,K_2$, the correlation functions $g_1(\epsilon,m)=\langle S_i S_{i+1} \rangle$,
$g_2(\epsilon,m)=\langle S_i S_{i+2} \rangle$, and $t(\epsilon,m)= \langle S_i S_{i+1} S_{i+2} \rangle$.

After this exposition of the reasons that make the method presented in in section \ref{sec:1bis} much more preferable with respect to a
direct counting, already for a simple Hamiltonian like (\ref{HAM12}), in the next subsection we give the results obtainable with our method
for this Hamiltonian. We find it useful to begin with the simple case with $K_2=0$, and then to proceed with the results obtained
for selected values of $K_2 \neq 0$. The comparison with the direct counting evaluation
will also be shown, although, as anticipated above, for $K_2 \neq 0$ the comparison is deferred to Appendix C.

\subsection{Results for the microcanonical entropy $\widetilde{s}_{\rm micr}(\epsilon,m)$}

\begin{figure}
  \centering
\includegraphics[width=0.65\textwidth]{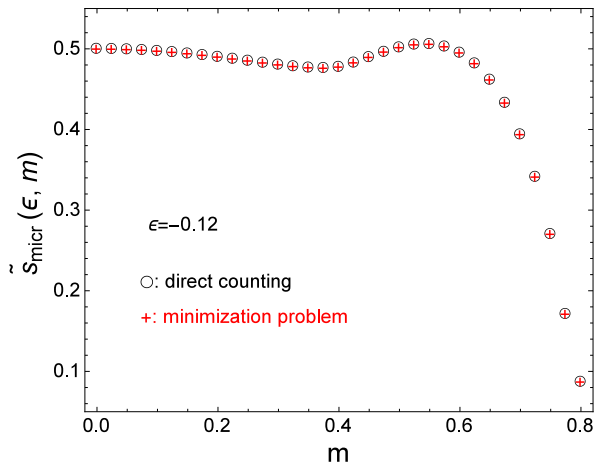}
\caption{ $\widetilde{s}_{\rm micr}(\epsilon,m)$ vs. $m$
  at the fixed value of the energy $\epsilon = -0.12$ (in units of $J=1$) for
  $K_1=-0.4$ and $K_2=0$. We plot the results from the minimization
  procedure discussed in sections \ref{sec:1} and \ref{sec:1bis} and from the direct counting
\cite{Mukamel05} (see as well subsection \ref{K2_0}).}
\label{fig1}
\end{figure}

As a first benchmark, in Fig. \ref{fig1} we plot the microcanonical entropy for the model with $K_2=0$ discussed in subsection \ref{K2_0}. The
comparison of the method presented in section \ref{sec:1bis} with the findings obtained by direct counting confirms the validity
of our results. Notice that in this section for simplicity we will set $J=1$. Indeed, our interest as previously discussed is on the
case $J \neq 0$ and moreover we remind that the model (\ref{HAM12}) does not exhibit magnetic order at finite temperature if $J$ is
negative (see \cite{Campa19} and references therein).

\begin{figure}
  \centering
\includegraphics[width=0.65\textwidth]{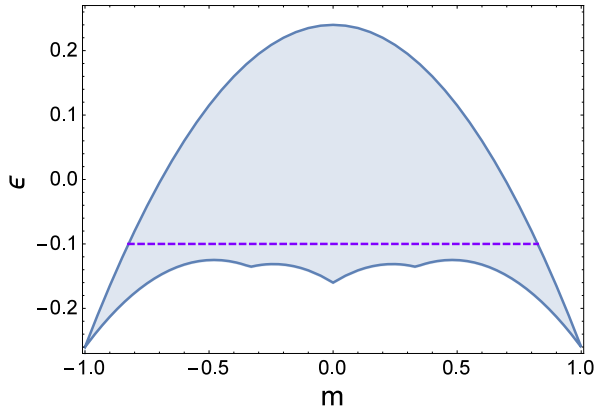}
\caption{Accessible region in the $m-\epsilon$ plane for $K_1=-0.4$ and $K_2=-0.08$. Blue dashed line shows the value of
the energy at which the first-order transition takes place ($\epsilon \simeq -0.100$).}
\label{fig2}
\end{figure}

\begin{figure}[!h]
\begin{center}
\begin{tabular}{ccc}
\includegraphics[width=0.485\textwidth]{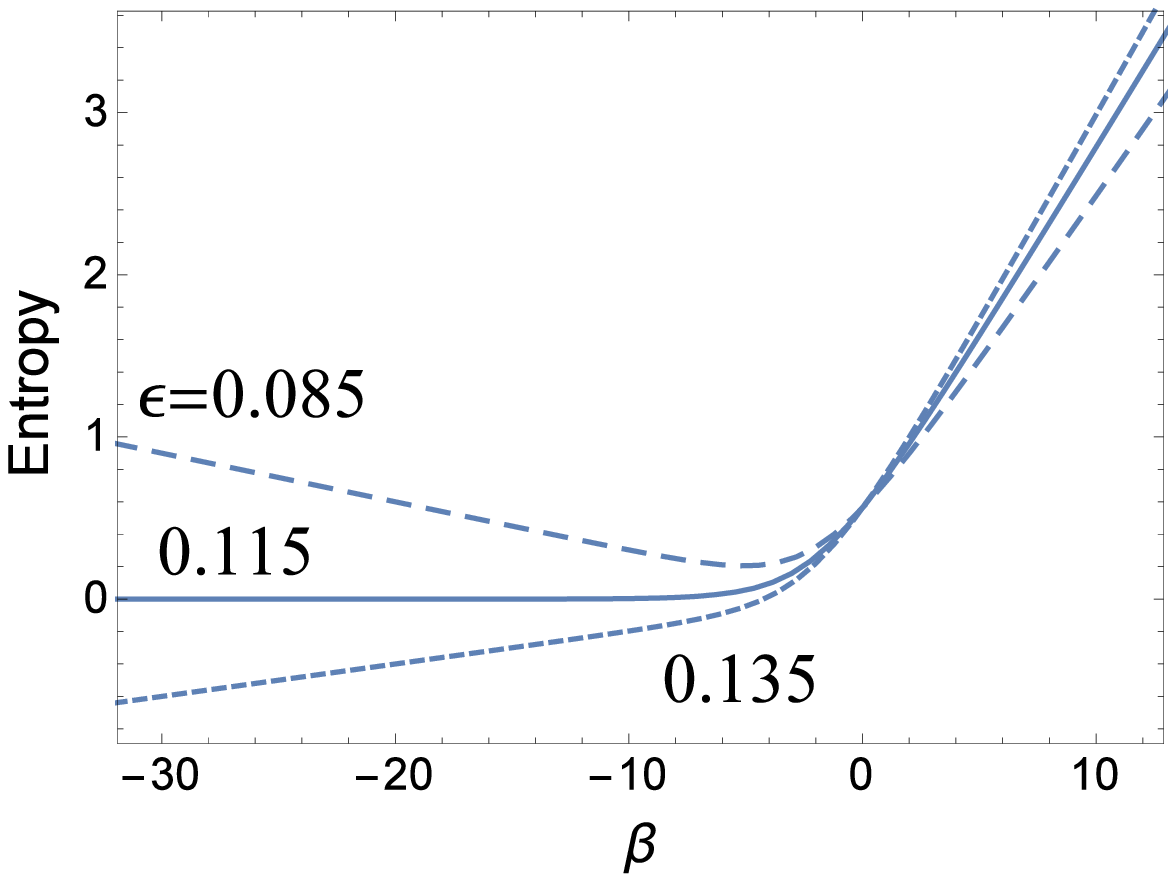} &
\includegraphics[width=0.5\textwidth]{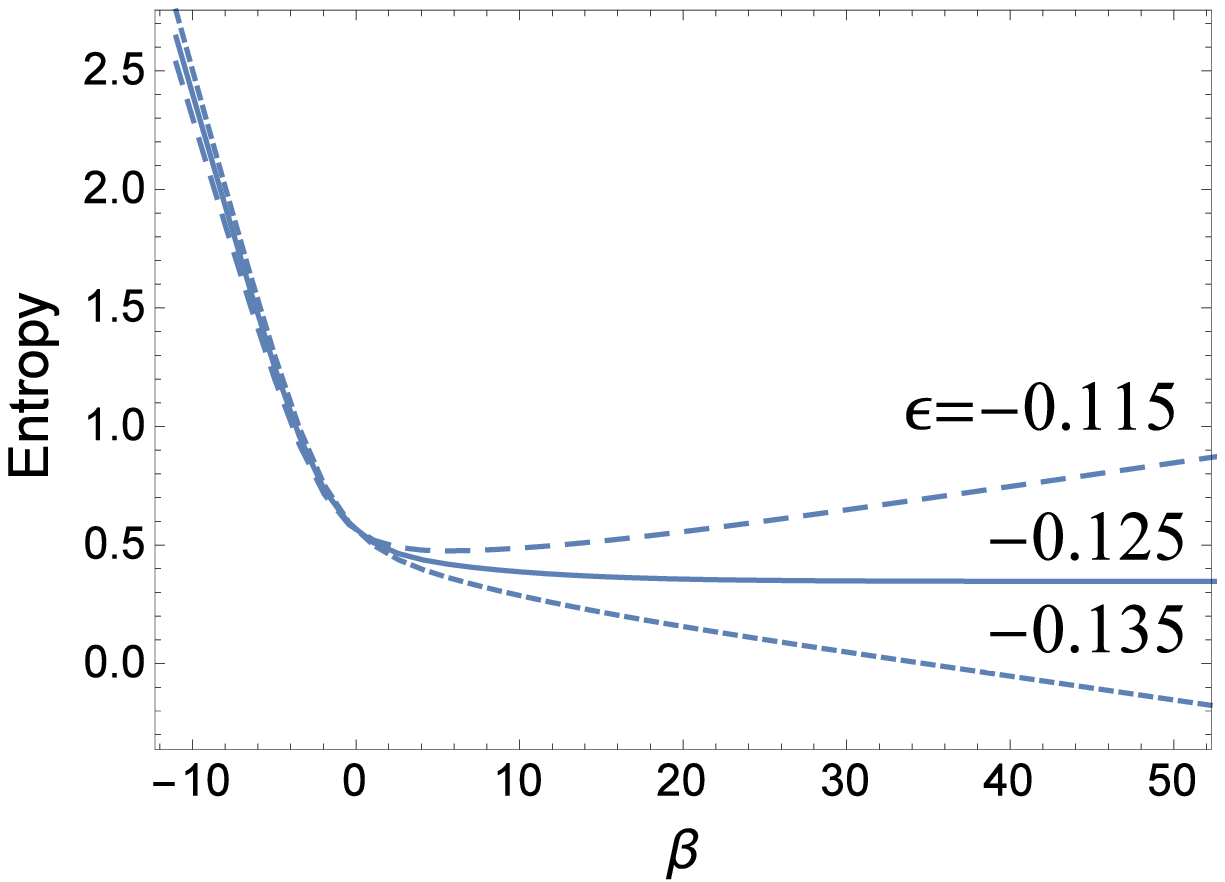}
\end{tabular}
\caption {Entropy from Eq. (\ref{micrentr_one}) as a function of $\beta$ after minimizing only with respect to $\varphi$ at
$K_1=-0.4$, $K_2=-0.08$ and $m=0.5$ for different values of $\epsilon$. The left and right panels show that for values of the energy
outside the allowed range for this value of $m$ (see Fig. \ref{fig2}), the entropy function has no minimum as a function
of $\beta$.}
\label{fig3}
\end{center}
\end{figure}

Let us pass now to the model with $K_2 \neq 0$. First, we remind that as a consequence of the nonadditivity of systems with long-range
interactions, intermediate values of the extensive variables may be not accessible. So a first important information is to determine the
accessible region in the $m-\epsilon$ plane. This is done in Fig. \ref{fig2} for fixed values of $K_1$ and $K_2$, chosen to be $K_1=-0.4$,
$K_2=-0.08$. Note that this pair of values corresponds in the $K_1$-$K_2$ plane to a point in a region which, despite the presence of a
non-vanishing $K_2$, has a phase diagram in the canonical ensemble in the space $K_1-T$ at fixed $K_2$ qualitatively similar to the phase
diagram in the space $K_1-T$ with $K_2=0$. Since for $K_1=-0.4$, $K_2=-0.08$ one knows that in the canonical ensemble there is a first
order phase transition line, as one can see from Fig. 2(left) of \cite{Campa19}, we also plot in Fig. \ref{fig2} the value of the energy
at which the first-order phase transition occurs. More details on the behaviour of the entropy function at fixed magnetization after the
minimization on the variable $\varphi$ are given in Fig. \ref{fig3}, where the same values of $K_1$, $K_2$ of Fig. \ref{fig2} are chosen.
As seen in Fig. \ref{fig2}, for $m=0.5$ the possible energy values are in the range $[-0.125, 0.115]$. Correspondingly, in Fig. \ref{fig3}
we see that for energy values outside this range, i.e., 0.135 in the left panel and -0.135 in the right panel, the entropy function
(precisely, the function of $\beta$ obtained from Eq. (\ref{micrentr_one}) after minimizing only with respect to $\varphi$) does
not have a minimum as a function of $\beta$, i.e., the microcanonical  extremal problem has no solution.

\begin{figure}
  \centering
\includegraphics[width=0.65\textwidth]{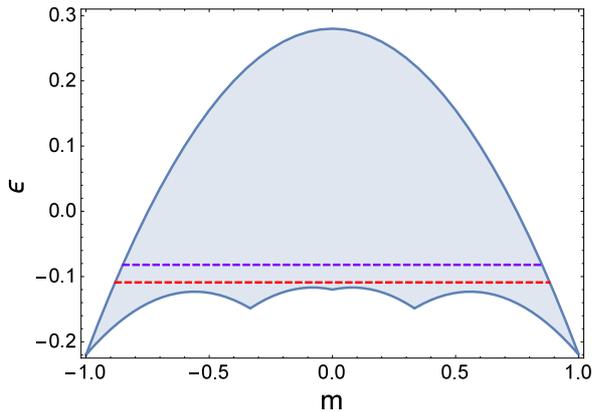}
\caption{Accessible region in the $m-\epsilon$ plane at $K_1=-0.4$ and $K_2=-0.16$. Red (blue) dashed lines show the value of the
energies at which the first-order (second-order) transition takes place: they are given respectively by $\epsilon \simeq -0.109$
($\epsilon \simeq -0.082$).}
\label{fig4}
\end{figure}

\begin{figure}
\includegraphics[width=0.95\textwidth]{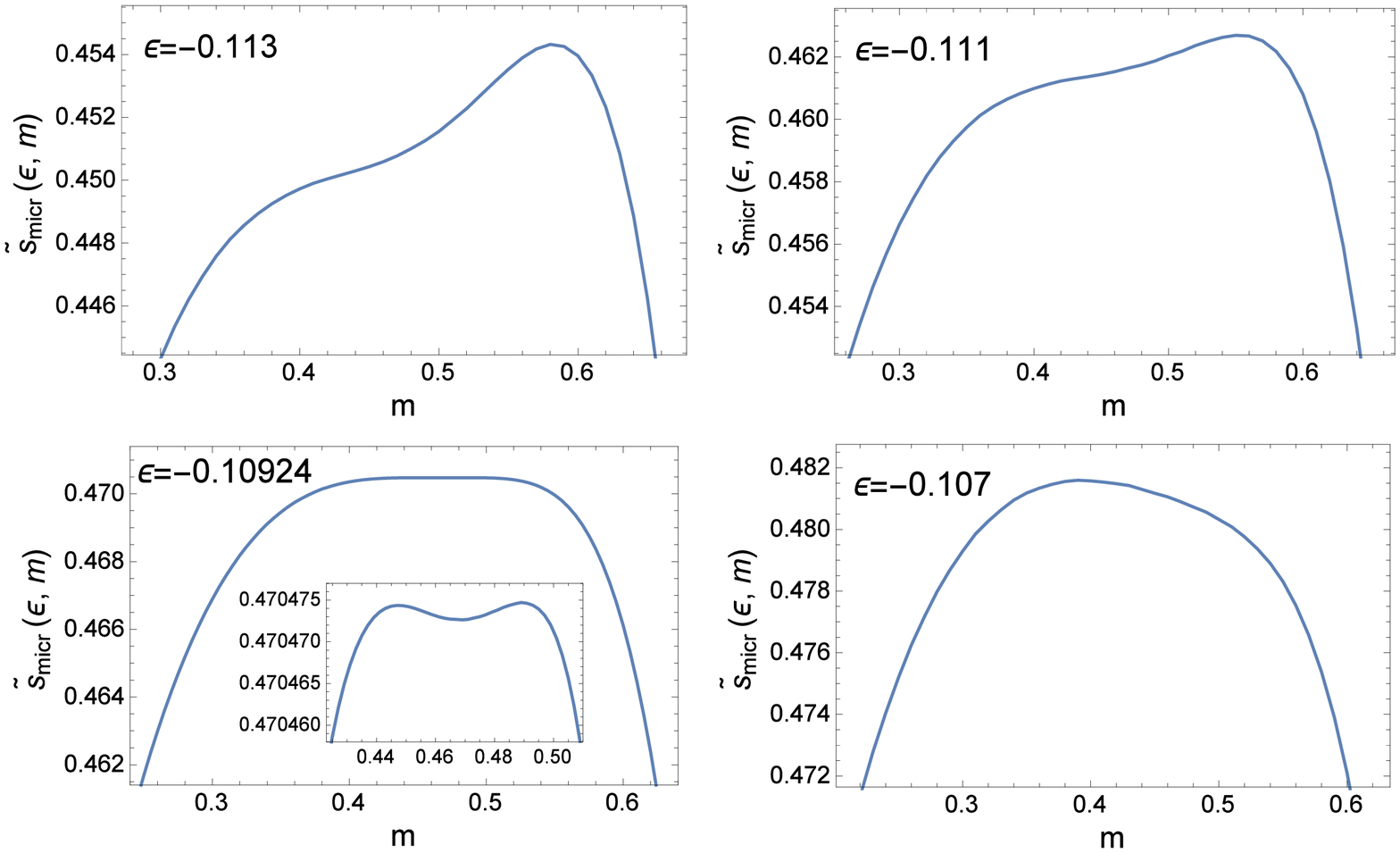}
\caption{$\widetilde{s}_{\rm micr}(\epsilon,m)$ vs. $m$ at the fixed value of the $K_1=-0.4$, $K_2=-0.16$ and few different values
of $\epsilon$ near the first-order transition.}
\label{fig5}
\end{figure}

\begin{figure}
\includegraphics[width=0.95\textwidth]{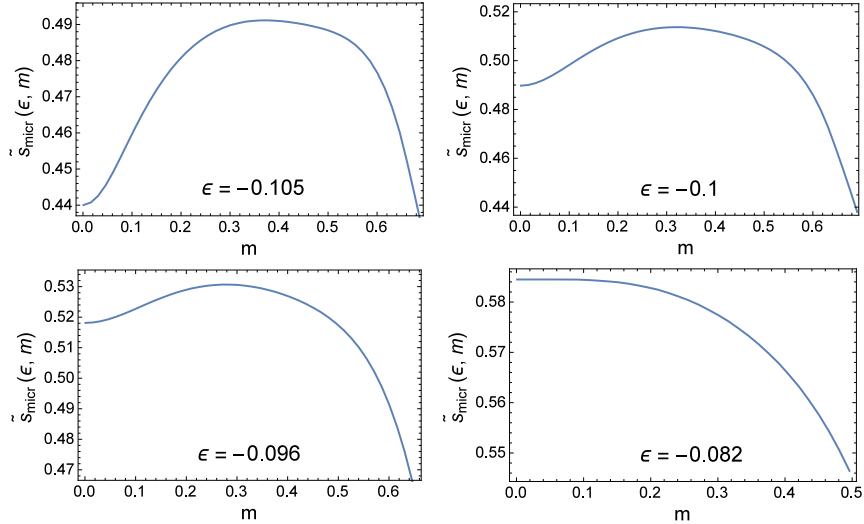}
\caption{$\widetilde{s}(\epsilon,m)$ vs. $m$ at the fixed value of the $K_1=-0.4$, $K_2=-0.16$ and few different values
of $\epsilon$ near the second-order transition.}
\label{fig6}
\end{figure}

To illustrate the effectiveness of our method, let us consider a pair of values of $K_1$, $K_2$ for which the model (\ref{HAM12}) is
known to have both a first-order and a second-order phase transitions in the canonical phase diagram. For this reason we consider
$K_1=-0.4$, $K_2=-0.16$. As one can see from Fig. 3(left) of \cite{Campa19}, when the temperature $T$ is increased at these particular
values of $K_1$, $K_2$, one meets a first-order transition at a certain temperature, and then at a larger temperature a second-order
transition. It is then very interesting to see what happens in the microcanonical ensemble when the energy is varied. Our main results
are summarized in Figs.  \ref{fig4}-\ref{fig5}-\ref{fig6}. In Fig. \ref{fig4} we plot the accessible region in the $m-\epsilon$ plane,
and we mark the energies at which the first- and second-order transitions occur. One observes that the shape of the accessible region
acquires further structure at low energies. Details on the behaviour of the microcanonical entropy near the first- and second-order
transitions are given respectively in Fig. \ref{fig5} and \ref{fig6}, from which one sees that the proposed method could be used to work
out the phase diagram in the microcanonical ensemble. In order to have a comparison, in Appendix C we provide an explicit example of the
determination of the microcanonical entropy at fixed magnetization corresponding to a particular point in Fig. \ref{fig5} using direct
counting, i.e. the expression of the entropy as a function of $(m,g_1,g_2)$ for a given energy. Agreement is found with the results
of Fig. \ref{fig5}.

\section{Conclusions}
\label{concl}
In this paper we presented a method to determine the microcanonical entropy at fixed magnetization starting from the canonical
partition function. We applied our results to the case of systems having long- and short-range (possibly competing) interactions. The
rationale behind this choice is that for models with only short-range interactions the canonical and microcanonical entropies, and in
particular the canonical and microcanonical entropies at fixed magnetization, do coincide, while this is not the case for models with
long-range interactions, as our construction explicitly shows. We also discussed in the Appendix A the connection with large-deviation
theory.

The presented method is based on the introduction of one (or more) auxiliary variables and on a min-max procedure, where the minimization
is performed on the variable $\beta$, which can be both positive or negative. We emphasized that the method can be very useful where
direct counting is not applicable or very difficult/convoluted.

We studied a model in which there is a long-range, all-to-all term in the presence of nearest-neighbour ($K_1$) and next-nearest-neighbour
($K_2$) couplings. Results for the microcanonical entropy at fixed magnetization of this model were presented, including a case in which
the canonical phase diagram exhibits first- and second-order phase transitions. The discussion clearly shows that increasing the range
of couplings (or including multi-spin interactions), even though an expression for the entropy in terms of all possible couplings can be
derived, the determination of the microcanonical entropy at fixed magnetization by direct counting requires the maximization over a number
of variables increasing with the range of the short-range interaction, while the method presented in section \ref{sec:1bis} -- once that
one has determined the partition function, which can be done by determining the largest eigenvalue of the transfer matrix -- requires the
extremization on a given number of variables equal to one plus the number of auxiliary variables, which, e.g., is just one for the model
(\ref{HAM12}), independently form the range of the short-range interaction. A discussion of advantages and disadvantages of the presented
method with the direct counting has been provided and a comparison with direct counting both with $K_2=0$ and $K_2 \neq 0$ for illustration
purposes has also been presented.

In the considered model the long-range interaction is of mean-field form, and it would be interesting as a future work to study the model
in which the interaction decays in space as a power-law. One could also consider more complicated short-range terms, such as involving more
than two spin interactions or couplings between spins up to a finite general $R$ larger than $2$. Moreover, our results show that the
presented scheme can be used to determine the phase diagram in the microcanonical ensemble, and a deserving application would be to work out
in detail the microcanonical phase diagram of Hamiltonian (\ref{HAM12}) in the whole $K_1-K_2$ space, and compare it with the corresponding
results in the canonical ensemble determined in \cite{Campa19}.

\begin{acknowledgements}
Discussions with N. Defenu, D. Mukamel and N. Ananikian are gratefully acknowledged. A.C. acknowledges financial support from INFN
(Istituto Nazionale di Fisica Nucleare) through the projects DYNSYSMATH and ENESMA.
G.G. is supported  by  the  Deutsche  Forschungsgemeinschaft (DFG, German Research Foundation) under Germany's Excellence
Strategy EXC 2181/1 - 390900948 (the Heidelberg STRUCTURES Excellence Cluster).
The authors acknowledge support by the RA MES Science
Committee and National Research Council of the Republic of Italy in the frames of the joint research project No. SCS 19IT-008 and
``Statistical Physics of Classical and Quantum Non Local Hamiltonians: Phase Diagrams and Renormalization Group''. This work
is part of MUR-PRIN2017 project ``Coarse-grained description for non- equilibrium systems and transport phenomena (CO-NEST)''
No. 201798CZL whose partial financial support is acknowledged.
\end{acknowledgements}

\section*{Appendix A: Derivation using large deviation techniques}
\label{append_large}

The basic expressions of this paper, i.e., Eqs. (\ref{exphitwo}), (\ref{canentrtwo}) and (\ref{micrentrtwo}), or equivalently
Eqs. (\ref{exphitwoboth}), (\ref{canentrtwoboth}) and (\ref{micrentrtwoboth}), for the rescaled free energy and the canonical and
microcanonical entropies, respectively, have been obtained starting from the formal expressions (\ref{micrpartspin}) and
(\ref{canpartspin}). The latter have been adapted to the models with Hamiltonian of the type (\ref{modhamext}) obtaining the
expressions (\ref{canpart0two}) and (\ref{micrpart0two}). In this Appendix we show how one can arrive at the basic expressions by
using an approach based on large deviation techniques. Of course, since one arrives at the same basic expressions for the rescaled
free energy and the entropies, the analysis presented in section \ref{sec:1bis} remains identical. Here, for brevity, we show the
procedure by starting directly from models with Hamiltonian of the type (\ref{modhamext}), therefore, without writing the more general
expressions and then adapting them to that kind of Hamiltonian.

Using the definitions (\ref{magdef}) and (\ref{quaddef}) of the magnetization $\hat{m}$ and the quadrupole moment $\hat{q}$,
respectively, and introducing the definition
\begin{equation}
\hat{r} = \frac{1}{N}\sum_{i=1}^N U([S_i]) \,
\end{equation}
the Hamiltonian (\ref{modhamext}) can be written as
\begin{equation}
H(\{S_i\}) = N \left[ -\frac{J}{2} \hat{m}^2 -\frac{K}{2}  \hat{q}^2 + \hat{r} \right] \, .
\label{modhamextlarge}
\end{equation}
At this point one formally defines
\begin{eqnarray}
\label{entrlarge}
&&\widehat{\Omega}(m,q,r,N) \equiv \exp \left[ N \widehat{s}_{{\rm micr}}(m,q,r) \right] = \\
&=& \sum_{\{S_i\}} \delta \left( N\hat{m} -Nm \right) \, \delta \left( N\hat{q} -Nq \right) \,
\delta \left( N\hat{r} -Nr \right) \, . \nonumber
\end{eqnarray}
One also defines the following kind of partition function
\begin{eqnarray}
\label{canpartlarge}
&&\widehat{Z}(\lambda_1,\lambda_2,\lambda_3,N) \equiv \exp \left[ - N \widehat{\phi} (\lambda_1,\lambda_2,\lambda_3) \right] =
\\
&=& \sum_{\{S_i\}} \exp \left[ -\lambda_1 \hat{m} -\lambda_2 \hat{q} -\lambda_3 \hat{r} \right] \, . \nonumber
\end{eqnarray}
In analogy with what we have noted for $\beta$ in the main text, the fact that the energy is upper bounded allows to consider both
signs for the parameters $\lambda_1$, $\lambda_2$ and $\lambda_3$. It is not difficult to see that in the thermodynamic limit the
function $\widehat{s}_{{\rm micr}}(m,q,r)$ and $\widehat{\phi}(\lambda_1,\lambda_2,\lambda_3)$ are related by the Legendre-Fenchel
transformation
\begin{equation}
\widehat{\phi}(\lambda_1,\lambda_2,\lambda_3) =
\min_{m,q,r} \left[ \lambda_1 m + \lambda_2 q + \lambda_3 r - \widehat{s}_{{\rm micr}}(m,q,r) \right] \, .
\label{lgtranslarge}
\end{equation}
In principle this transformation is not invertible. However, if $\widehat{\phi}$ is everywhere differentiable, the inversion is possible,
so to have
\begin{equation}
\widehat{s}_{{\rm micr}}(m,q,r) =  \min_{\lambda_1,\lambda_2,\lambda_3}
\left[ \lambda_1 m + \lambda_2 q + \lambda_3 r - \widehat{\phi}(\lambda_1,\lambda_2,\lambda_3) \right] \, .
\label{invlgtranslarge}
\end{equation}
It can be seen, in analogy with the function $\hat{\psi}$ defined in Eq. (\ref{psihatdeftwo}), that the function
$\widehat{\phi}(\lambda_1,\lambda_2,\lambda_3)$ is differentiable (basically, this is assured from the fact that the expressions in
the exponent in the right hand side of Eq. (\ref{canpartlarge}) are sums of one-particle functions). Then,
Eq. (\ref{invlgtranslarge}) is verified. Once $\widehat{s}_{{\rm micr}}(m,q,r)$ is given, our function of interest,
$\widetilde{s}_{{\rm micr}}(\epsilon,m)$ is obtained from
\begin{equation}
\label{micrentrlarge}
\widetilde{s}_{{\rm micr}}(\epsilon,m) =
\max_{\left[ q,r | - \frac{J}{2}m^2 - \frac{K}{2} q^2 + r = \epsilon \right]} \widehat{s}_{{\rm micr}}(m,q,r) \, .
\end{equation}

It remains to see that from the last expression we can derive Eq. (\ref{micrentrtwoboth}). Substituting Eq. (\ref{invlgtranslarge}) and
expressing $r$ as a function of $\epsilon$, $m$ and $q$, we have
\begin{eqnarray}
&&\widetilde{s}_{{\rm micr}}(\epsilon,m) = \\
&=& \max_q \left\{ \min_{\lambda_1,\lambda_2,\lambda_3}
\left[ \lambda_1 m + \lambda_2 q + \lambda_3 \left( \epsilon + \frac{J}{2}m^2 + \frac{K}{2}q^2\right)
- \widehat{\phi}(\lambda_1,\lambda_2,\lambda_3) \right] \right\} \, . \nonumber
\label{micrentrlarge_b}
\end{eqnarray}
The four stationarity conditions of this problem are:
\begin{eqnarray}
\label{condlarge1}
m - \frac{\partial \widehat{\phi}}{\partial \lambda_1} &=& 0 \\
\label{condlarge2}
q - \frac{\partial \widehat{\phi}}{\partial \lambda_2} &=& 0 \\
\label{condlarge3}
\epsilon + \frac{J}{2}m^2 + \frac{K}{2}q^2 - \frac{\partial \widehat{\phi}}{\partial \lambda_3} &=& 0 \\
\label{condlarge4}
\lambda_2 + \lambda_3 Kq &=& 0 \, .
\end{eqnarray}
Eliminating $\lambda_2$ from the problem by implementing immediately the last stationarity condition, we have
\begin{eqnarray}
&&\widetilde{s}_{{\rm micr}}(\epsilon,m) = \\
&=& \max_q \left\{ \min_{\lambda_1,\lambda_3}
\left[ \lambda_1 m + \lambda_3 \left( \epsilon + \frac{J}{2}m^2 - \frac{K}{2}q^2\right)
- \widehat{\phi}(\lambda_1,-\lambda_3 Kq,\lambda_3) \right] \right\} \, . \nonumber
\label{micrentrlarge_c}
\end{eqnarray}
After the notation changes $\lambda_3 \to \beta$ and $q \to y$, and defining $\varphi$ by $\lambda_1 = - \beta J m -\varphi$, the
last expression becomes
\begin{eqnarray}
&&\widetilde{s}_{{\rm micr}}(\epsilon,m) = \\
&=& \max_y \left\{ \min_{\beta,\varphi}
\left[ \lambda_1 m + \beta \epsilon -\beta \frac{J}{2}m^2 - \beta \frac{K}{2}q^2
- \widehat{\phi}(-\beta Jm -\varphi,-\beta K y,\beta) \right] \right\} \, . \nonumber
\label{micrentrlarge_d}
\end{eqnarray}
This is recognized to be the same as Eq. (\ref{micrentrtwoboth}), once one realizes from the definitions (\ref{psihatdeftwo})
and (\ref{canpartlarge}) that $\widehat{\phi}(-\beta Jm -\varphi,-\beta K y,\beta) = \hat{\psi}(\beta,\beta J m + \varphi,\beta K y)$.

\section*{Appendix B: General expressions for the entropy obtained with direct counting}
\label{append_gen}

In this Appendix we provide a summary of the general expressions, obtained with direct counting, for the entropy of one-dimensional Ising spin
models having the form $H=H_{LR}+H_{SR}$, where $H_{LR}=-\frac{J}{2N}\left(\sum_i S_i \right)^2$ is the long-range, mean-field term, and
$H_{SR}$ is the short-range part of the Hamiltonian. The interested reader can find full details in Ref. \cite{Gori11}. We emphasize
that we are referring to the expressions of the entropy defined as the logarithm of the number of configurations for given values of
the magnetization and of the spin correlations, like, e.g., the function $s(m,g_1,g_2,t)$ in (\ref{entropy2}). From these expressions
one can obtain the microcanonical entropy $\widetilde{s}_{\rm micr}(\epsilon,m)$ with the long optimization procedure
described in subsection 4.1.

In the main text we confined ourselves to the form (\ref{HSR}), where only two-spin terms are included. The short-range Hamiltonian
(\ref{HSR}) is a sub-case of the general one-dimensional short-range Ising model with multispin interactions defined by
\begin{displaymath}
  H_{SR}=-\sum_{i} j^{(1)}_{i} S_{i} - \sum_{i,j} j^{(2)}_{i,j} S_{i} S_{j} -
  \sum_{i,j,k} j^{(3)}_{i,j,k} S_{i} S_{j} S_{k}
\end{displaymath}
\begin{equation}
- \sum_{i,j,k,l} j^{(4)}_{i,j,k,l} S_{i} S_{j} S_{k} S_{l} - \ldots
\label{HamiltonianExplicit}
\end{equation}
where the sums run over distinct couples, triples, quartets and so on up to a certain finite range. Periodic boundary conditions are assumed
and the couplings $j^{(n)}$ are assumed to be invariant under translation by $\rho$ spins:
\begin{equation}
 j^{(n)}_{i_1,i_2,\ldots,i_n}= j^{(n)}_{i_1+\rho,i_2+\rho,\ldots,i_n+\rho} .
\end{equation}
As in the main text, $N$ denotes the number of sites and $R$ the finite-range of the interaction. For example, the Hamiltonian
$H_{SR}=-(K_1/2) \sum_i S_i S_{i+1}$ has $\rho=1$ and $R=2$, while $H_{SR}=-(K_1/2) \sum_i S_i S_{i+1}-(K_2/2) \sum_i S_i S_{i+2}$
has $\rho=1$ and $R=3$. In the general case, for simplicity we assume $N/\rho$ is an integer.

Following the notation and the procedure presented in \cite{Gori11}, let us start from the case $J=0$, for which $H=H_{SR}$. To simplify
the notation let us rewrite (\ref{HamiltonianExplicit}) as
\begin{equation}
 H_{SR} \equiv - \sum_{\mathrm{Rg}(\mu)\le R}' \sum_{n=1}^{N/\rho}
j_{\mu} O_{\mu+n \rho} (\{S_i\}) \label{Hamiltonian} \, ,
\end{equation}
where $\mu$ is a subset of $\{1,\ldots, R\}$. The notation $\mathrm{Rg}(\mu)\le R$, stands for ``the range of the interaction is less
than or equal to $R$''. Moreover, $O_{\mu+n \rho}$ is an operator associated to the subset
$\mu \equiv \{n_{1}, n_{2}, \ldots n_{|\mu|}\}$, where $|\mu|$ is the number of elements of $\mu$, and translated by
$i \rho$ so that it acts on the spins as
\begin{equation}
 O_{\mu+i \rho} (\{S_i\}) =S_{n_{1} + n \rho} S_{n_{2} + n \rho} \ldots
S_{n_{|\mu|} + n \rho}.
\end{equation}
For the null subset $\varnothing$ we define $O_{\varnothing}=1$ and the prime $'$ in the sum over $\mu$ in (\ref{Hamiltonian}) denotes
that the null subset is not included and that the terms related by a translation of a multiple of $\rho$ are counted only once. The
correlation functions are denoted by $g_{\mu}$. They are associated to the operator $O_{\mu}$ and defined according to
\begin{equation}
  g_{\mu}=\langle O_{\mu}(\{S_i\})\rangle = \langle S_{n_{1}} S_{n_{2}} \ldots S_{n_{|\mu|}} \rangle \,
  \label{corr}
\end{equation}
(by definition, $g_{\varnothing}=1$). For example, for the Hamiltonian considered in subsection 4.1 we would have the correlation
functions $g_{\{1\}}$, $g_{\{1,2\}}$, $g_{\{1,3\}}$ and $g_{\{1,2,3\}}$, that in the lighter notation of the main text were denoted,
respectively, with $m$, $g_1$, $g_2$ and $t$. 

The main result of Ref. \cite{Gori11} concerns the entropy $s(\{g_{\mu}\})$ for the model with interactions up to
range $R$; it is given by: 
\begin{equation}
 s(\{g_{\mu}\})=s^{(R)}(\{g_{\mu}\})-s^{(R-\rho)}(\{g_{\mu}\}) \, ,
\label{entropy}
\end{equation}
where in the left side and in the first term in the right hand side $\{g_{\mu}\}$ stands for the set of all possible
correlations of range up to $R$, while in the seond term in the right hand side it stands for all correlations
of range up to $(R-\rho)$. 
It is written in terms of the functions $s^{(Q)}(\{g_{\mu}\})$. The quantity $s^{(Q)}$ can be seen as
the ``entropy at range $Q$'', and is given by
\begin{equation}
 s^{(Q)}(\{g_{\mu}\})=-\sum_{\tau_{Q}} p(\tau_{Q}) \ln p( \tau_{Q}) \, ,
 \label{satrange1}
\end{equation}
where
\begin{equation}
p(\tau_{Q})=2^{-Q}\sum_{\mathrm{Rg}(\mu)\le Q} g_{\mu} O_{\mu} (
\tau_{Q}) \, , \label{satrange2}
\end{equation}
and $\tau_Q \equiv \{t_1, t_2, \ldots, t_Q\}$ denotes the configuration of $Q$ Ising spins, with the sum over $\mu$ is on every
subset (including the null one). Finally, from (\ref{Hamiltonian}) one gets the energy per unit cell as:
\begin{equation}
 e(\{g_{\mu}\})=-\sum_{1 \le \mathrm{Rg}(\mu)\le R}
g_{\mu} j_{\mu} \, .
\end{equation}

When the mean-field term $H_{LR}$ is turned on, only the energy $e$ is affected, while the dependence of the entropy
$s(\{g_{\mu}\})$ on the correlations $g_{\mu}$ is not. One then has to add the corresponding contribution to $e$. In this way one finds
the results (\ref{entropy1}) and (\ref{en1}), respectively for $s$ and for $e$, for the model (\ref{HAM12}) with $K_2=0$; and the results
(\ref{entropy2}) and (\ref{en2}) for the same model with $K_2 \neq 0$. In particular, specializing (\ref{entropy}) to our
model with $K_2=0$ one has that Eq. (\ref{entropy1}) is obtained from
$s(m,g_1) = s^{(2)}(m,g_1) - s^{(1)}(m)$, while Eq. (\ref{entropy2}) for the model with $K_2 \neq 0$ is obtained from
$s(m,g_1,g_2,t) = s^{(3)}(m,g_1,g_2,t) - s^{(2)}(m,g_1)$.  

\section*{Appendix C: Comparison of the microcanonical entropy at fixed magnetization for $K_2 \neq 0$}

In this Appendix we consider an example of explicit determination of the microcanonical entropy at fixed magnetization directly from
the entropy $s=s(m,g_1,g_2,t)$, given  in Eq. (\ref{entropy2}), for the model (\ref{HAM12}). To compare the findings with the results
obtained with the method presented in section \ref{sec:1bis}, we choose the same values of $K_1$ and $K_2$ used in Fig. \ref{fig5}:
$K_1=-0.4$, $K_2=-0.16$ (with $J=1$). The energy is chosen as $\epsilon=-0.107$, as in the bottom right panel of Fig. \ref{fig5}.

As discussed in the main text, one has to determine $t$ via Eqs. (\ref{derivata})-(\ref{cub}). Once this is done, one has to express $g_2$
as a function of $m,g_1$ using the energy expression (\ref{en2}). One has then $s$ as a function of $m$ and $g_1$ and it is possible
to plot in the $m-g_1$ plane the allowed regions. Of course the same procedure can be performed by studying the entropy in the $m-g_2$
plane. The final point in both cases is to find the maximum of the microcanonical entropy maximizing with respect to, respectively,
$g_1$ or $g_2$.

\begin{figure}
\includegraphics[width=0.48\textwidth]{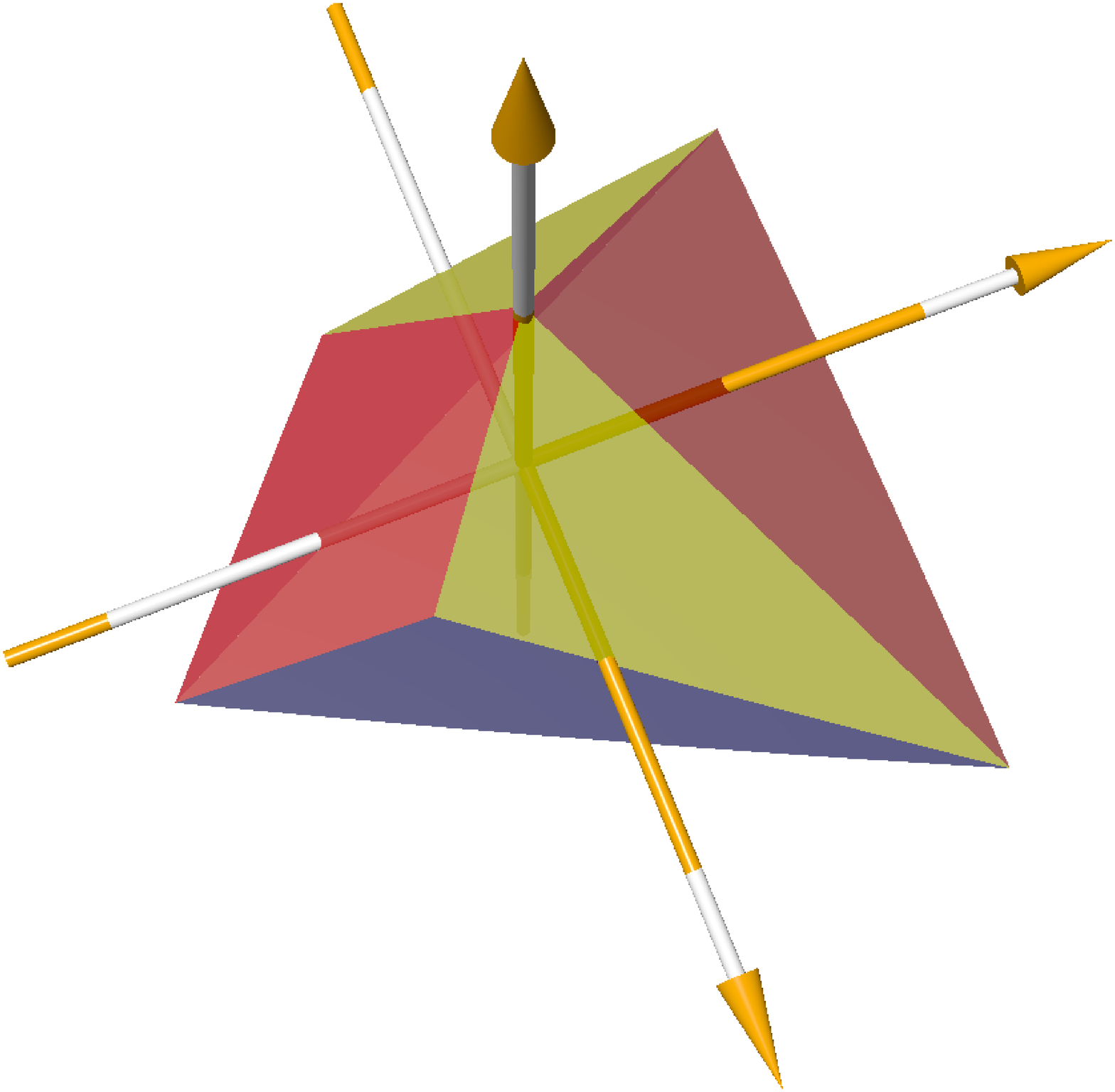}
\begin{picture}(0,0)
\put(-38,72){$g_1$}
\put(-78,0){$m$}
\put(-78,110){$-g_2$}
\end{picture}
\includegraphics[width=0.48\textwidth]{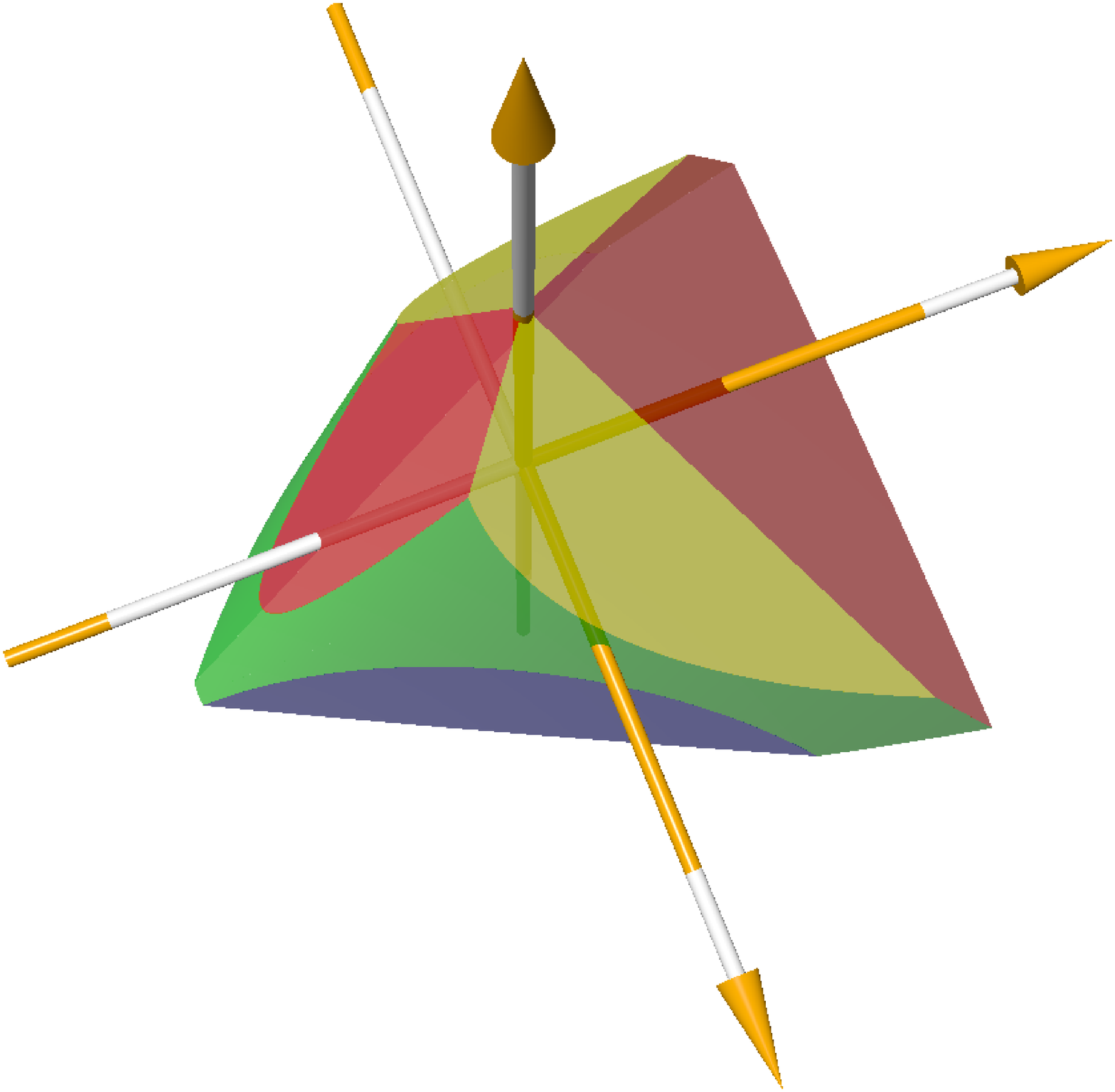}
\begin{picture}(0,0)
\put(-38,72){$g_1$}
\put(-78,0){$m$}
\put(-78,110){$-g_2$}
\end{picture}
\caption{Three-dimensional plot of the allowed region in the $(m,g_1,g_2)$ space (left). On the right the allowed region has been cut
with the constant energy surface (a parabolic cylinder) given by $\epsilon = -\frac{1}{2}(J m^2+K_1 g_1+K_2 g_2)$.
The green surface is thus the accessible region in the microcanonical ensemble. The chosen values are $\epsilon=-0.107$, $K_1=-0.4$
and $K_2=-0.16$. Please note that in order to improve visibility the $g_2$ axis has been reversed.}
\label{fig1app}
\end{figure}

\begin{figure}
\includegraphics[width=0.49\textwidth]{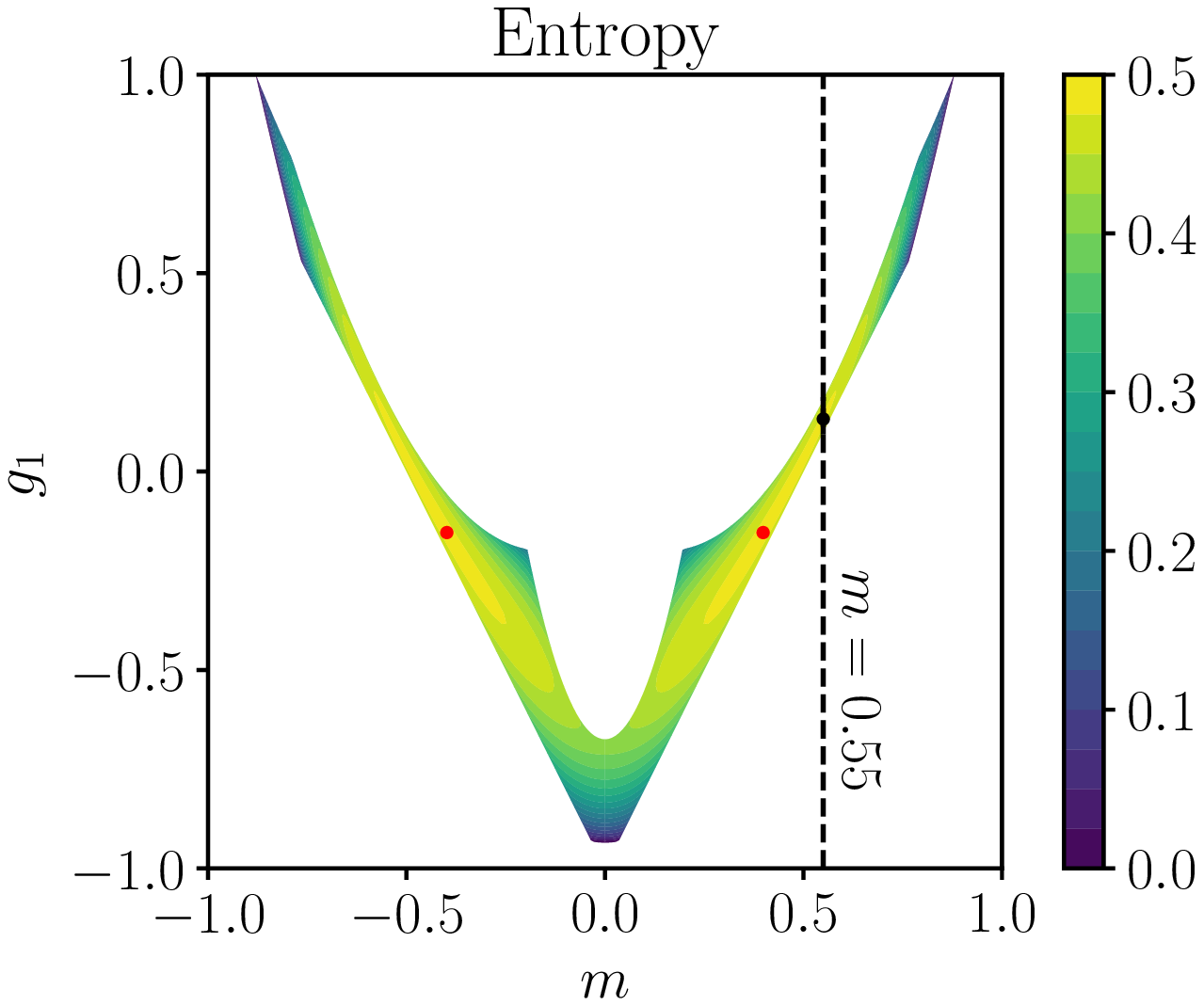}
\includegraphics[width=0.49\textwidth]{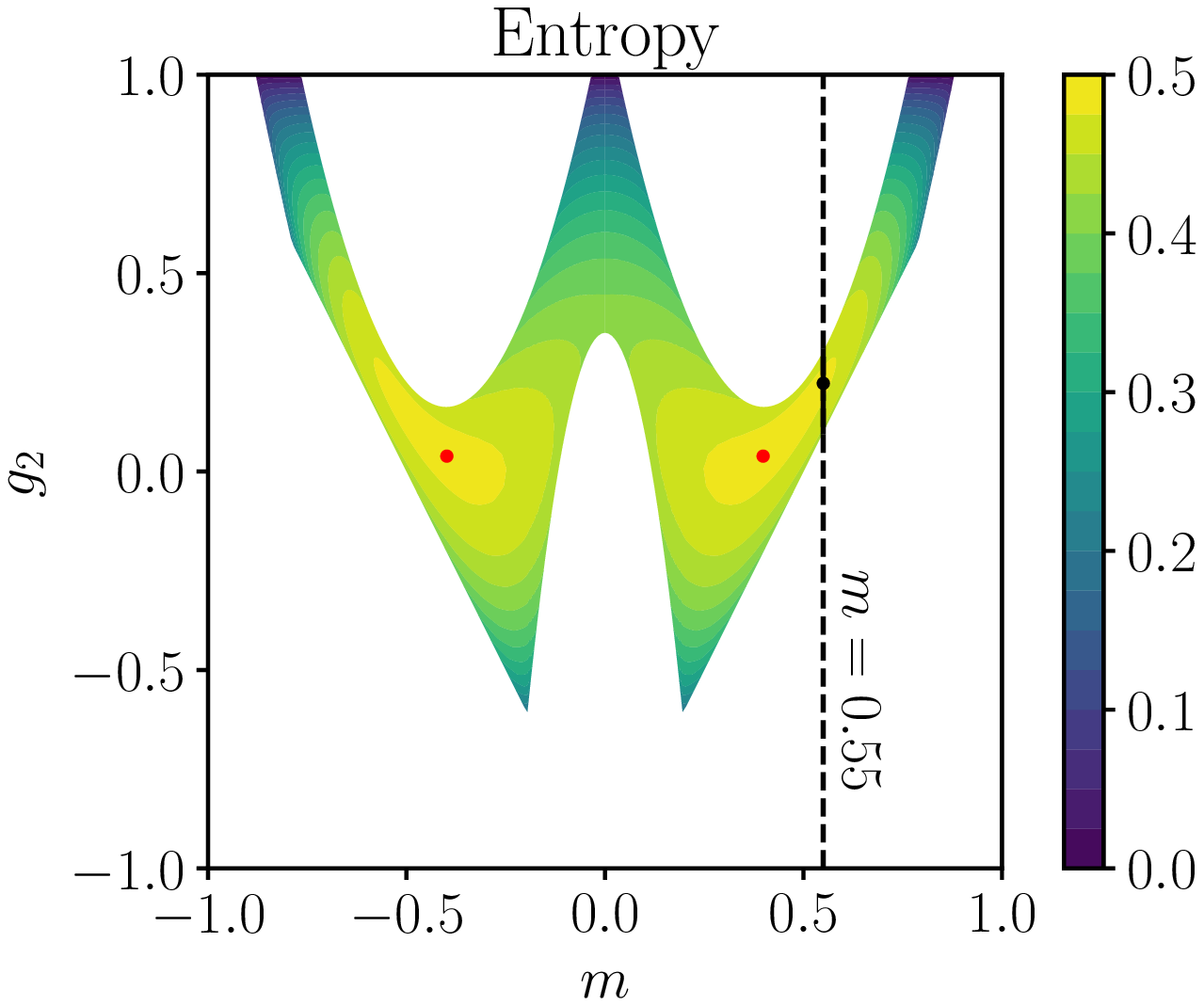}
\caption{Entropy as a function of $m$ and $g_1$ (left) and $m$ and $g_2$ (right) for the parameter values $\epsilon=-0.107$,
$K_1=-0.4$ and $K_2=-0.16$. The line with magnetization $m=0.55$ (chosen to have an example of comparison with Fig. \ref{fig5}) is denoted
with a black line. The maximum of the entropy $s$ in this fixed magnetization sector is denoted with a black dot. It is characterised by
the observables $m=0.55$, $g_1 \simeq 0.13235$, $g_2 \simeq 0.22226$ and $t \simeq -0.16384$. The global maxima of $s$ are also shown with
a red dot. These points are characterized by the following observables: $m \simeq \pm 0.39839$, $g_1 \simeq -0.15375$,
$g_2 \simeq 0.03885$, and $t\simeq -0.46475$.}
\label{fig2app}
\end{figure}

\begin{figure}
\centering
\includegraphics[width=0.65\textwidth]{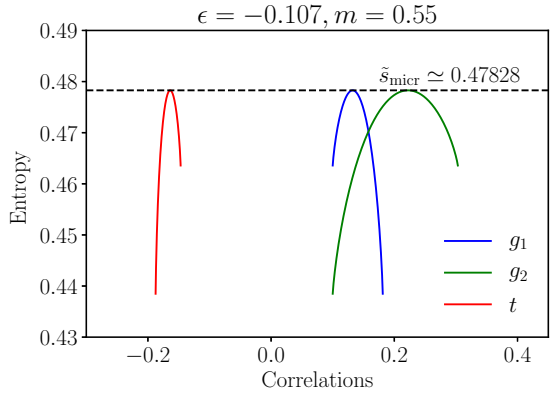}
\caption{Entropy in the fixed energy ($\epsilon=-0.107$) and magnetization ($m=0.55$) sector as a function of the three independent
correlations $g_1$ (blue), $g_2$ (green) and $t$ (red) when the other two are eliminated. Parameter values are $K_1=-0.4$ and $K_2=-0.16$.
The maximum  $\widetilde{s}_\mathrm{micr}$ is denoted with a dashed line and it occurs at $g_1 \simeq 0.13235$, $g_2 \simeq 0.22226$,
and $t\simeq -0.16384$. The value of the maximum, $\widetilde{s}_\mathrm{micr} \simeq 0.47828$, corresponds to the value
of $\widetilde{s}_{{\rm micr}}(\epsilon,m)$ at $m=0.55$ in the right bottom panel of Fig. \ref{fig5}.}
\label{fig3app}
\end{figure}

Notice that in this procedure finding the maximum with respect to $g_1$, $g_2$ and $t$ is the easier part since the entropy is concave
along these directions on the constant energy surface. In the remaining variable $m$ instead, within the constant energy surface, the
entropy is {\it not} concave and many entropy maxima can and do appear and compete resulting in the emergence of the different phases and
transitions among them. This is obviously to be traced to the special role of $m$ in the Hamiltonian, in which it appears nonlinearly and thus
can spontaneously break the $m\rightarrow -m$ symmetry. Restricting to fixed magnetization indeed relieves many of the
difficulties. The procedure is described in Figs. \ref{fig1app}-\ref{fig2app}, where we consider the value $m=0.55$, and one finds
that the maximum entropy is $s\simeq 0.47828$, in agreement with the results presented in the bottom right panel of Fig. \ref{fig5}, obtained
with the procedure described in section \ref{sec:1bis}. Note that in the course of the process we also determine the macroscopic observables fully
characterizing the thermodynamic state. Fig. \ref{fig3app} also shows that if we decide to eliminate $g_2$ in favour of $(m,g_1)$ or,
alternatively,  $g_1$ in favour of $(m,g_2)$, we obtain the same result for the microcanonical entropy at fixed magnetization when the maximum in,
respectively, $g_1$ or $g_2$ is taken, as of course it has to be. For completeness we also plot the entropy as a function of the correlation $t$
after maximizing with respect to one among $g_1$ and $g_2$.

The direct counting method outlined shows some technical difficulties due to the already moderately large number of variables over which
the entropy has to be optimized. Of course these extra variables are interesting in their own right being macroscopic observables fully
characterising the thermodynamic state. On the other hand direct counting possesses the virtue of making very clear the geometric origin
of (microcanonical) phase transitions in long-range systems as the study of the maxima of the entropy restricted on the nonlinear energy
surface. This makes interesting also short-range one-dimensional systems, whose entropy is concave in all variables, yielding normally no
phase transition. The gained insight could prove useful in the understanding of and the hunt for the many exotic critical points expected
in the microcanonical ensemble \cite{Barre05}.

\section*{Conflict of interest}

The authors declare that they have no conflict of interest.


\begin{thebibliography}{}
\bibitem{Huang}
K. Huang, Statistical mechanics. Wiley, New York (1987).
\bibitem{Pathria}
R. K. Pathria, Statistical mechanics. Butterworth-Heinemann, Oxford (1996).
\bibitem{PR}
A. Campa, T. Dauxois, and S. Ruffo, Phys. Rep. {\bf 480}, 57 (2009).
\bibitem{Parisi}
G. Parisi, Statistical field theory. Wiley, New York (1987).
\bibitem{Misawa}
T. Misawa, Y. Yamaji, and M. Imada, J. Phys. Soc. Jpn. {\bf 75},  064705 (2006).
\bibitem{Auerbach}
A. Auerbach, Interacting electrons and quantum magnetism.  Springer-Verlag, New York (1994).
\bibitem{Matsubara}
T. Matsubara and H. Matsuda, Prog. Theor. Phys. {\bf 16}, 416 (1956).
\bibitem{Lacroix}
Introduction to frustrated magnetism: materials, experiments, theory, eds. C. Lacroix, P. Mendels,
and F. Mila (Heidelberg, Springer, 2011).
\bibitem{Nagle70}
J. F. Nagle, Phys. Rev. A, {\bf 2}, 2124 (1970).
\bibitem{Kardar83}
M. Kardar, Phys. Rev. B, {\bf 28}, 244 (1983).
\bibitem{Mukamel05}
D. Mukamel, S. Ruffo, and N. Schreiber, Phys. Rev. Lett., {\bf 95}, 240604 (2005).
\bibitem{Campa19}
A. Campa, G. Gori, V. Hovhannisyan, S. Ruffo, and A. Trombettoni,
J. Phys. A: Math. Theor. {\bf 52}, 344002 (2019).
\bibitem{campa04}
A. Campa and A. Giansanti, Physica A {\bf 340}, 170 (2004).
\bibitem{leyv2002}
F. Leyvraz and S. Ruffo, J. Phys. A: Math. Gen. {\bf 35}, 285 (2002).
%\bibitem{Campa14}
%A. Campa, T. Dauxois, D. Fanelli and S. Ruffo,
%Physics of long-range interacting systems. Oxford University Press, Oxford (2014).
\bibitem{Ruelle}
D. Ruelle, Statistical mechanics: rigorous results. Benjamin, New York, (1969).
\bibitem{BMR2001}
J. Barr\'e, D. Mukamel, and S. Ruffo, Phys. Rev. Lett. {\bf 87}, 030601 (2001).
\bibitem{Gori11}
G. Gori and A. Trombettoni, J. Stat. Mech., P10021 (2011).
\bibitem{Barre05}
F. Bouchet and J. Barr\'e, J. Stat. Phys. {\bf 118}, 1073 (2005).
\end{thebibliography}
\end{document}